\documentclass[conference]{llncs}
\usepackage[cmex10]{amsmath}
\usepackage{amssymb}

\newenvironment{mmproof}{\hspace{8pt}\ti{Proof:}}{}

\newcommand{\ie}{\emph{i.e.}, }
\newcommand{\eg}{\emph{e.g.}, }
\newcommand{\pnt}[1]{{\mbox{\boldmath $#1$}}}
\newcommand{\cof}[2]{\mbox{$#1_{\boldsymbol{#2}}$}}
\newcommand{\cf}[3]{\mbox{$#1^{#2}_{\boldsymbol{#3}}$}}

\newcommand{\V}[1]{\mbox{$\mathit{Vars}(#1)$}}
\newcommand{\Va}[1]{\mbox{$\mathit{Vars}(\boldsymbol{#1})$}}

\newcommand{\s}[1]{\mbox{$\{#1\}$}}

\newcommand{\nGz}[2]{$G_{non-\{z\}}$}

\newcommand{\prr}[1]{\mi{Prev}(\boldsymbol{q})}

\newcommand{\mi}[1]{\mathit{#1}}
\newcommand{\ti}[1]{\textit{#1}}
\newcommand{\tb}[1]{\textbf{#1}}

\newcommand{\Ds}[2]{\mbox{\pnt{#1}~$\rightarrow$ \s{#2}}}
\newcommand{\Dds}[2]{\mbox{\pnt{#1}~$\rightarrow #2$}~}

\newcommand{\Tt}{\>\>}
\newcommand{\Sub}[2]{\mbox{$#1_\mi{#2}$}}
\newcommand{\Sup}[2]{\mbox{$#1^\mi{#2}$}}
\newcommand{\Dii}{\Sub{\mi{DDS}}{impl}~}
\newcommand{\Di}{$\mi{DDS}~$}
\newcommand{\DI}{\mbox{$\mi{DDS}$}}
\newcommand{\Ac}[2]{$#1_{(#2)}$}
\newcommand{\Act}[3]{$#1^{#2}_{(#3)}$}

\newcommand{\Ss}[1]{\scriptsize{#1}}
\newcommand{\prob}[2]{\mbox{$\exists{#1} [#2]$}}
\newcommand{\DS}{\mbox{$\Omega$}}

\newcommand{\ecnf}{\ensuremath{\exists\mathrm{CNF}}}
\newcommand{\Xr}{\mbox{$X^{\Omega}$}}
\newcommand{\Xrr}[1]{\mbox{$X^{#1}$}}
\newcommand{\ax}{\mbox{{\boldmath $a^{\Omega}$}}}
\newcommand{\axx}[1]{\mbox{\boldmath $a^{#1}$}}

\newcommand{\Dss}[5]{\mbox{$(\prob{#1}{#2},\pnt{#3},#4)~\rightarrow #5$}}
\newcommand{\Ods}[4]{\mbox{$(\prob{#1}{#2},\pnt{#3})~\rightarrow #4$}}
\newcommand{\DDs}[5]{\mbox{$(\prob{#1}{#2},\pnt{#3},#4)~\rightarrow$ \s{#5}}}
\newcommand{\scp}{\mbox{$W^{\DS}$}}
\newcommand{\scpp}[1]{\mbox{$W^{#1}$}}

\usepackage{wrapfig}
\usepackage{graphicx}

\def \FullVersion {}

\begin{document}

\title{Quantifier Elimination by Dependency Sequents}

\author{Eugene Goldberg,  Panagiotis Manolios}
\institute{Northeastern University, USA \email{\{eigold,pete\}@ccs.neu.edu}}

\maketitle

\begin{abstract}
We consider the problem of existential quantifier elimination
for Boolean formulas in Conjunctive Normal Form (CNF).  We
present a new method for solving this problem called Derivation of
Dependency-Sequents (DDS). A Dependency-sequent (D-sequent) is
used to record that a set of quantified variables is redundant
under a partial assignment. We introduce a resolution-like operation called join
that produces a new D-sequent from two existing D-sequents. We also show that
DDS is compositional, \eg if our input formula is a conjunction
of independent formulas, DDS automatically recognizes and
exploits this information. We introduce an algorithm based on DDS
and present experimental results demonstrating its potential.
\end{abstract}

\section{Introduction}
In this paper, we consider the problem of eliminating existential
quantifiers from Boolean CNF formulas.  In the sequel, we omit
the word ``existential.''  Given a Boolean CNF formula
\prob{X}{F}, the problem is to find a quantifier-free  CNF formula
$G$ such that $G \equiv \prob{X}{F}$.  We assume that the set of non-quantified variables 
$\V{F} \setminus X$  is, in general, not empty. (\V{F} is the set of variables of $F$). So $G$ specifies a Boolean
function depending on non-quantified variables of $F$.
We refer
to this problem as the \tb{QE problem}, where QE stands for
Quantifier Elimination.

Our interest in the QE problem is twofold. First, the QE problem occurs in numerous 
areas of hardware/software design and
verification, \eg in symbolic model checking~\cite{mc,mc_thesis}
when computing reachable states. Second, one can argue that
progress in solving the QE problem should have a deep impact on
SAT-solving~\cite{HVC}. In particular, as McMillan pointed out,
even the basic operation of resolution is related to the QE
problem~\cite{blocking_clause}. The resolvent $C$ of clauses
$C'$,$C''$ on a variable $v$ is obtained by eliminating the
quantifier from $\exists{v}[C' \wedge C'']$.

The success of resolution-based SAT-solvers~\cite{grasp,chaff}
has led to the hunt for efficient SAT-based algorithms for the QE
problem~\cite{blocking_clause,cav09,cav11,cofactoring}. In this
paper, we continue in this direction by introducing a resolution-based
QE algorithm. Our approach is based on the following
observation. The QE problem is trivial if $F$ does not depend on variables of $X$.
 In this case, dropping the quantifiers from
\prob{X}{F} does not affect the meaning of the formula.  If
$F$ depends on $X$, after adding to $F$ a set of clauses implied
by $F$, the variables of $X$ may become redundant.  If this happens,
all the clauses of $F$ depending on $X$ can be dropped and the
resulting formula $G$ is equivalent to the original formula
\prob{X}{F}. The problem is that \ti{one needs to
  know when the variables of} $X$ \ti{become redundant}.

Unfortunately, resolution is deficient in expressing redundancy
of variables. Let $Y$ denote the set of non-quantified variables in \prob{X}{F} i.e. $Y = \V{F} \setminus X$.
 Let \pnt{y} be a complete assignment for
$Y$ and  \cof{F}{y} denote $F$ under assignment \pnt{y}.  Then a clause
$C$ falsified by \pnt{y} can be derived by resolving clauses of
$F$. After adding $C$ to $F$, the variables of $X$ are
redundant in $\prob{X}{\cof{F}{y}}$.  In this case, resolution
works. Assume, however, that $\cof{F}{y}$ is \ti{satisfiable}.
Then, the variables of $X$ are \ti{also redundant} in
$\prob{X}{\cof{F}{y}}$ because $\cof{F}{y}$ remains
satisfiable after removing any clauses. But a resolution
derivation cannot express this fact because no clause falsified
by \pnt{y} is implied by $F$.

To address the problem above, we introduce the notion of Dependency
sequents (\ti{D-sequents}). A D-sequent has the form
\Ods{X}{F}{q}{Z} where \pnt{q} is a partial assignment to variables of
$F$ and $Z \subseteq X$. This D-sequent states that in
the subspace specified by \pnt{q}, the variables of $Z$ are
redundant in \prob{X}{F}. That is in this subspace, the clauses
containing variables of $Z$ can be removed from $F$ without
changing the meaning of \prob{X}{F}. In particular, if
the formula \cof{F}{y} is satisfiable, the D-sequent \Ods{X}{F}{y}{X}
holds. For the sake of simplicity, 
in the introduction, we drop the parameter of scope used in the definition of D-sequents given in Section~\ref{sec:d_sequents}.

In this paper, we introduce a QE algorithm called \Di (Derivation
of D-Sequents). In \Di\!\!, adding resolvent clauses to $F$ is
accompanied by computing D-sequents. The latter are used to
\ti{precisely identify the moment when the variables of} $X$
\ti{are redundant}. It occurs when the D-sequent $(\prob{X}{F},\emptyset)
\rightarrow X$ is derived stating unconditional redundancy of
$X$.  Then, a solution $G$ to the QE problem is obtained
from $F$ by dropping the clauses containing variables of $X$.

\Di produces new D-sequents
from existing ones by operation \ti{join}.  Let  $(\prob{X}{F},\pnt{q_1}) \rightarrow Z$ and  $(\prob{X}{F},\pnt{q_2}) \rightarrow Z$ be 
 D-sequents where \pnt{q_1} and \pnt{q_2} have opposite
assignments to exactly one variable $v$. Then  a new  D-sequent
$(\prob{X}{F},\pnt{q})\!\!\rightarrow~\!\!Z$ can be obtained by joining the D-sequents above, where \pnt{q} contains all  assignments of
\pnt{q_1} and \pnt{q_2} but those to $v$. 

In this paper, we compare \Di with its counterparts both
theoretically and experimentally. In particular, we show that \Di
is \ti{compositional} while algorithms based on enumeration of
satisfying
assignments~\cite{blocking_clause,fabio,cofactoring,cav11} are
not.  Compositionality here means that given   formula
\prob{X}{F_1 \wedge \dots \wedge F_k} where formulas  $F_i$ depend on non-overlapping
sets of variables, \Di breaks the QE problem
into $k$ independent subproblems.  \Di is a branching algorithm
and yet it remains compositional no matter how branching
variables are chosen.  
Compositionality of \Di means that its performance can be
\ti{exponentially better} than that of enumeration-based QE
algorithms.  Since \Di is a branching algorithm it can process
variables of different branches in different orders.  This gives
\Di a big edge over QE algorithms that eliminate quantified
variables one by one using a global
order~\cite{cav09,HVC}.

D-sequents are tightly related to boundary points~\cite{bnd_pnts}.  A
boundary point is a complete assignment to variables of $F$ with
certain properties.  To make variables of $Z \subseteq X$
redundant in \prob{X}{F}, one needs to eliminate a particular set of
boundary points. This elimination is performed by adding to $F$
resolvent clauses that do not depend on variables of $Z$.  \Di
does not compute boundary points \ti{explicitly}. 
 Nevertheless, we introduce them
in this paper because boundary points
provide the semantics of \Di\!\!\!. In particular, 
the notion of  scoped variable redundancy  we use in this paper can be formulated 
only in terms of boundary points.

The contribution of this paper is as follows. First, we relate
the notion of variable redundancy with the elimination of boundary
points.  Second, we introduce the notion of D-sequents and the
operation of joining  D-sequents. Third, we introduce \DI, a QE
algorithm; we prove its correctness and evaluate it 
experimentally.  Fourth, we show that \Di is compositional.

This paper is structured as follows. In Section~\ref{sec:rvars_bps}, we relate the 
notions of variable
redundancy and boundary points. Section~\ref{sec:div_conq}
explains the strategy of \Di in terms of boundary point
elimination.  Two simple cases of variable redundancy are described  in Section~\ref{sec:trivial_cases} 
and D-sequents are introduced in
Section~\ref{sec:d_sequents}.  Sections~\ref{sec:alg_descr} and
~\ref{sec:compos} describe \Di and discuss its compositionality.
Section~\ref{sec:experiments} gives experimental results. 
Background is discussed in Section~\ref{sec:background}, and
conclusions are presented in Section~\ref{sec:conclusion}. In the appendix,
we describe some details of the implementation of DDS we used in experiments
and give proofs of propositions.

\section{Redundant Variables, Boundary Points and Quantifier Elimination}
The main objective of this section is to introduce the notion of
redundant variables (Definition~\ref{def:red_vars}) and to relate it
to the elimination of removable boundary points
(Proposition~\ref{prop:red_vars}).
\label{sec:rvars_bps}
%
%
\subsection{Redundant Variables and Quantifier Elimination}
In this paper, we consider a quantified CNF formula
$\prob{X}{F}$ where $X \subseteq \V{F}$ 
 We will refer to such formulas as {\boldmath $\exists$}\tb{CNF}.
 Let \pnt{q} be an
assignment, $F$ be a CNF formula, and $C$ be a clause. \Va{q}
denotes the variables assigned in \pnt{q}; \V{F} denotes the set
of variables of $F$; \V{C} denotes the variables of $C$;
and $\V{\prob{X}{F}} = \V{F} \setminus X$.

%
%
\begin{definition}
\label{def:cofactor}
Let $C$ be a clause, $F$ be a CNF formula, and \pnt{p} be an 
assignment such that $\Va{p} \subseteq \V{F}$. \cof{C}{p} is \tb{true} if $C$ is satisfied by
\pnt{p}; otherwise it is the clause obtained from $C$ by removing
all literals falsified by \pnt{p}.  \cof{F}{p} denotes the
CNF formula obtained from $F$ 
by  replacing every clause $C$ with \cof{C}{p} and then removing all the clauses that are true (i.e. satisfied by \pnt{p}).
If $\V{F}
\subseteq \Va{p}$, then $\cof{F}{p}$ is semantically equivalent
to a constant, and in the sequel, we will make use of this
without explicit mention.
\end{definition}
%
%
\begin{definition}
\label{def:ecnf_cofactor}
Let \prob{X}{F} be an \ecnf formula  and \pnt{p} be an 
assignment such that $\Va{p} \subseteq \V{\prob{X}{F}}$.  Denote by  \cof{(\prob{X}{F})}{p} the \ecnf formula
\prob{X}{\cof{F}{p}}. If $\V{\prob{X}{F}}
\subseteq \Va{p}$, $\Va{p} \cap X = \emptyset$, then $\cof{(\prob{X}{F})}{p}$ is semantically equivalent
to a constant, and in the sequel, we will make use of this
without explicit mention.
\end{definition}

%
%
\begin{definition}
\label{def:qe-solution}
The Quantifier Elimination (QE) problem for \ecnf{} formula
$\exists X [F]$ consists of finding a CNF formula $G$ such that
$G \equiv \exists X [F]$. This equivalence means that \cof{G}{p} = \cof{(\prob{X}{F})}{p} holds
for every complete assignment \pnt{p} to the variables
of $\V{G} \cup \V{\prob{X}{F}}$.
\end{definition}
%
%
\begin{definition}
\label{def:X_clause}
 A clause $C$ of $F$ is called a \pnt{Z}\tb{-clause} 
 if \V{C} $\cap~Z~\neq~\emptyset$. Denote by
           {\boldmath $F^{Z}$} the set of all $Z$-clauses of
           $F$.
\end{definition}
%
%
\begin{definition}
\label{def:red_vars}
The variables of $Z$ are 
\textbf{redundant} in CNF formula $F$ if 
$F \equiv (F \setminus F^Z)$.
The variables of $Z$ are 
\textbf{redundant} in \ecnf{} formula
$\exists X [F]$ if 
$\exists X [F] \equiv \exists X [F \setminus F^Z]$. We note that 
since $F \setminus F^Z$ does not contain any $Z$ variables, we
could have written  $\exists (X \setminus Z) [F \setminus
F^Z]$. To simplify notation, we avoid explicitly using this
optimization in the rest of the paper. 
\end{definition}

%
%
\subsection{Redundant Variables and Boundary Points}
%
%
\begin{definition}
\label{def:pnt}
Given assignment \pnt{p} and a formula $F$, we say that 
\pnt{p} is an $F$-\textbf{point} (or a \textbf{point} of $F$)
if $\V{F} \subseteq \Va{p}$.
\end{definition}

In the sequel, by ``assignment'' we mean  a possibly partial one. To refer to a \ti{complete}
assignment we will use term ``point''.
%
%
\begin{definition}
\label{def:bnd_pnt}
A point \pnt{p} of CNF formula $F$ is called a
\pnt{Z}\tb{-boundary point} of $F$ if a) $Z \neq \emptyset$, b)
$\cof{F}{p} = \mi{false}$; c) every clause of $F$ falsified by
\pnt{p} is a $Z$-clause; d) the previous condition breaks for
every proper subset of $Z$.
\end{definition}

Suppose that \pnt{p} is a $Z$-boundary point of $F$ and $F$ is
satisfiable. If only $Z$ variables can be flipped in
\pnt{p}, then it is at least $|Z|$ flips away from a satisfying
assignment, hence the name ``boundary.''

%
%
\begin{definition}
\label{def:rem_bnd_pnt}
Given a CNF formula $F$ and a $Z$-boundary point, \pnt{p}, of
$F$:
\begin{itemize} 
\item \pnt{p} is
$X$-\tb{removable} in $F$ if 1) $Z \subseteq X \subseteq \V{F}$; and 2) there is a
clause $C$ such that a) $F \Rightarrow C$; b) $\cof{C}{p} =
\mathit{false}$; and c) $\V{C} \cap X = \emptyset$. 

\item \pnt{p} is
\tb{removable} in $\prob{X}{F}$ if \pnt{p} is $X$-removable in $F$.
\end{itemize} 
\end{definition}

In the above definition, notice that \pnt{p} is not a
$Z$-boundary point of $F \wedge C$ because \pnt{p} falsifies $C$
and $\V{C}\cap Z = \emptyset$.
%
%
\begin{proposition}
\label{prop:rem_bnd_pnt}
A $Z$-boundary point \pnt{p} of $F$ 
is removable in \prob{X}{F},
iff one cannot turn \pnt{p} into an assignment satisfying $F$
by changing only the values of variables of $X$.
\end{proposition}

The proofs are given in the appendix of this paper.
%
%
\begin{proposition}
\label{prop:red_vars}
The variables of $Z \subseteq X$ are not redundant in \prob{X}{F} iff there is an  
$X$-removable $W$-boundary point of $F$, $W \subseteq Z$.
\end{proposition}

Proposition~\ref{prop:red_vars} justifies the following strategy
of solving the QE problem. Add to $F$ a set $G$ of clauses that
a) are implied by $F$; b) eliminate all $Z$-removable boundary
points for all $Z \subseteq X$. By dropping all $X$-clauses of $F$,
one produces a solution to the QE problem. 

Below we  introduce the notion of scoped redundancy of variables.
We use  the notion of scoped redundancy in the definition of dependency sequents (Section~\ref{sec:d_sequents}).
%
%
\begin{definition}
\label{def:scoped_red_vars}
Let $Z$ be a set of variables redundant in \prob{X}{F} where $Z \subseteq X$.
We will say that the variables of $Z$ are \tb{redundant} in \prob{X}{F} \tb{with scope} $W$ where $W \supseteq Z$ if
for any non-empty subset $V \subseteq Z$, the set of $W$-removable $V$-boundary points
is empty. In other words, any $V$-boundary point of $F$ where $V \subseteq Z$ can be turned into an assignment satisfying $F$ by flipping
only variables of $W$. We will say that the variables of  $Z$ are \tb{locally redundant} in \prob{X}{F} if the scope of their redundancy is 
equal to $Z$.
\end{definition}

Notice that if variables of $Z$ are redundant in \prob{X}{F} with scope $W$ they are also redundant in \prob{X}{F} in terms of Definition~\ref{def:red_vars}.
The opposite is not true. 
Informally, $W$ can be viewed as a measure of how  hard it is to prove redundancy of $Z$.
The larger $W$, the harder the proof. The notion of scoped redundancy  is used in this paper 
instead of  that of virtual redundancy\footnote{\label{fnote:var_red}
In~\cite{qe_tech_rep3}, we used the notion of virtual redundancy to address the following problem. The fact that 
$\prob{X}{\cof{F}{s}} \equiv \prob{X}{\cof{F}{s} \setminus (\cof{F}{s})^Z}$
does not imply that  $\prob{X}{\cof{F}{q}} \equiv \prob{X}{\cof{F}{q} \setminus (\cof{F}{q})^Z}$ where $\pnt{s} \subset \pnt{q}$.
That is redundancy of  variables $Z$  in subspace \pnt{s}  specified by Definition~\ref{def:red_vars}
 does not imply such redundancy in  subspace \pnt{q} contained
in subspace \pnt{s}. The notion of virtual redundancy solves this paradox by \tb{weakening} Definition~\ref{def:red_vars}. Namely,
variables of $Z$ are redundant in \pnt{q} even if $\prob{X}{\cof{F}{q}} \not\equiv \prob{X}{\cof{F}{q} \setminus (\cof{F}{q})^Z}$
but $\prob{X}{\cof{F}{s}} \equiv \prob{X}{\cof{F}{s} \setminus (\cof{F}{s})^Z}$ for some \pnt{s} such that $\pnt{s} \subset \pnt{q}$.
In this paper, we solve the problem above by using scoped redundancy i.e. by \tb{strengthening} Definition~\ref{def:red_vars}. The trick
is that  we forbid to assign variables of scope $W$. Then (see Lemma~\ref{lemma:holds_in_subspace} of the appendix),
redundancy of  $Z$  with scope $W$ in subspace \pnt{q} where  $W \cap \Va{s}=\emptyset$ implies redundancy of $Z$ in any subspace 
\pnt{q} where $\pnt{s} \subset \pnt{q}$ if $W \cap \Va{q} = \emptyset$.
}  introduced in the previous version of this paper~\cite{qe_tech_rep3}.

From now on, when we say that variables of $Z$ are redundant in \prob{X}{\cof{F}{q}} with scope $W$ we will assume
that $W \cap \Va{q} = \emptyset$.

\section{Boundary Points And Divide-And-Conquer Strategy} 
\label{sec:div_conq}
In this section, we provide the semantics of the QE algorithm \Di described in Section~\ref{sec:alg_descr}.
\Di is a branching algorithm. Given an \ecnf{} formula \prob{X}{F}, it branches on variables of $F$
 until proving redundancy of  variables of $X$ 
in the current subspace becomes trivial.
Then \Di merges the results obtained in different branches to prove that the variables of $X$ are redundant in the entire search space.
Below we give propositions justifying the divide-and-conquer strategy  of \Di\!\!. 
Proposition~\ref{prop:glob_impl_loc} shows how to perform elimination of removable boundary points
of $F$ in the subspace specified by assignment \pnt{q}. This is done by using formula \cof{F}{q}, a ``local version'' of $F$.
 Proposition~\ref{prop:rem_bpts_increm} justifies proving redundancy of variables of $X$ in \cof{F}{q} one by one.

%
%

Let \pnt{q} and \pnt{r} be  assignments to a set of variables $Z$.
Since \pnt{q} and \pnt{r} are sets of value assignments to individual variables of $Z$
one can apply set operations to them. We will denote by $\pnt{r} \subseteq \pnt{q}$
the fact that \pnt{q} contains all the assignments \pnt{r}.  The  assignment  consisting of value assignments of \pnt{q} and \pnt{r} is represented
as   $\pnt{q} \cup \pnt{r}$.

%
%
\begin{proposition}
\label{prop:glob_impl_loc}
Let \prob{X}{F} be an \ecnf{} formula and \pnt{q} be an  assignment to \V{F}. Let \pnt{p} be a  $Z$-boundary point 
of $F$ where $\pnt{q} \subseteq \pnt{p}$ and $Z \subseteq X$. Then if \pnt{p} is removable in \prob{X}{F} it is also removable
in \prob{X}{\cof{F}{q}}.
\end{proposition}

\begin{remark}
\label{rem:glob_impl_loc}
Proposition~\ref{prop:glob_impl_loc} is not true in the opposite direction.
That is, a boundary point  may be $X$-removable in \cof{F}{q} and not $X$-removable in $F$.
For instance, if $X=\V{F}$, a $Z$-boundary point \pnt{p} of $F$ is removable in \prob{X}{F} for any $Z \subseteq X$ 
only by adding an empty clause to $F$. So if $F$ is satisfiable,
\pnt{p} is not removable in \prob{X}{F}. Yet \pnt{p} may be removable in \prob{X}{\cof{F}{q}} if \cof{F}{q} is unsatisfiable.
\end{remark}

%
%
\begin{proposition}
\label{prop:rem_bpts_increm}
Let \prob{X}{F} be a CNF formula and \pnt{q} be an assignment to variables of $F$.
Let the variables of $Z$ be  redundant in \prob{X}{\cof{F}{q}} with scope $W$ where $Z \subseteq (X \setminus \Va{q})$. Let
a variable  $v$ of $X \setminus (\Va{q} \cup Z)$ be locally redundant in $\prob{X}{\cof{F}{q} \setminus (\cof{F}{q})^Z}$.
Then the variables of $Z \cup \s{v}$ are  redundant in \prob{X}{\cof{F}{q}} with scope $W \cup \s{v}$.
\end{proposition}

Proposition~\ref{prop:rem_bpts_increm} shows that one can prove redundancy of variables of $X \setminus \Va{q}$  \ti{incrementally},
if every \s{v}-clause is removed from \cof{F}{q} as soon as variable $v$ is proved redundant.

\section{Two Simple Cases of Local Variable Redundancy}
\label{sec:trivial_cases}
In this section, we describe two easily identifiable cases where variables
are locally redundant (see Definition~\ref{def:scoped_red_vars}). 
These cases are specified by Propositions~\ref{prop:bl_var_red} and~\ref{prop:unsat_clause}.
%
\begin{definition} Let $C'$ and $C''$ be  clauses having opposite
  literals of exactly one variable $v \in \V{C'} \cap \V{C''}$. 
The clause $C$ consisting of all literals of $C'$ and $C''$ but
those of $v$ is called the \tb{resolvent} of $C'$,$C''$ on $v$.
Clause $C$ is said to be obtained by \tb{resolution} on $v$.
Clauses $C'$,$C''$ are called \tb{resolvable} on $v$.
\end{definition}
%
%
\begin{definition}
\label{def:blocked_var}
 A variable $x$ of a CNF formula $F$ is called \tb{blocked} if  no two clauses of $F$ are resolvable on $x$. 
A  \tb{monotone} variable $x$ (literals of only one polarity of $x$ are
 present in $F$) is a special case of a blocked variable.
\end{definition}

The notion of blocked variables is related to that of blocked clauses introduced in~\cite{blocked_clause} (not to confuse
with \ti{blocking} clauses  ~\cite{blocking_clause}). A clause $C$ of $F$ is blocked with respect to  $x$ if
no clause $C'$ of $F$ is resolvable with $C$ on $x$. Variable $x$
is blocked in $F$ if every \s{x}-clause of $F$ is blocked with respect to $x$.
\begin{proposition}
\label{prop:bl_var_red}
Let \prob{X}{F} be an \ecnf{} formula and \pnt{q} be an assignment to \V{F}. Let a variable $v$ of  
 $X \setminus \Va{q}$ be
blocked in \cof{F}{q}. Then $v$ is locally redundant in \prob{X}{\cof{F}{q}}.
\end{proposition}
\begin{proposition}
\label{prop:unsat_clause}
Let \prob{X}{F} be an \ecnf{} formula and \pnt{q} be an assignment to \V{F}. 
Let \cof{F}{q} have an empty clause.
Then the variables of  $X \setminus \Va{q}$ are locally redundant in \prob{X}{\cof{F}{q}}.
\end{proposition}

\section{Dependency Sequents (D-sequents)}
\label{sec:d_sequents}

In this section, we define D-sequents and introduce the operation of joining D-sequents.
We also introduce the notion of composable D-sequents\footnote{As far as composability of D-sequents is concerned, we made two changes  in comparison to paper~\cite{qe_tech_rep3}. 
First, we use term 'composable' instead of 'mergeable'
and term 'compatible' instead of 'consistent'. Second, in~\cite{qe_tech_rep3} we put the discussion of composability of D-sequents
into the appendix. In the current paper, we split this discussion between the main body of the paper and the appendix.

}. 

%
\subsection{Definition of D-sequents}
\label{subsec:def_d_seqs}
%
%
\begin{definition}
\label{def:d_sequent}
Let \prob{X}{F} be an \ecnf{} formula. Let \pnt{q} be an assignment to \V{F}
and $Z$ be a subset of $X \setminus \Va{q}$. Let  $W$ be a set of variables such that  $Z \subseteq W \subseteq (X \setminus \Va{q})$.
A dependency sequent (\textbf{D-sequent})  has the form
\Dss{X}{F}{q}{W}{Z}. It states that  the variables of $Z$ are redundant in  \prob{X}{\cof{F}{q}} with scope $W$.
\end{definition}

The definition above is different from those given in previous versions of this paper~\cite{qe_tech_rep1,qe_tech_rep3}.
A brief discussion of this topic is given below\footnote{In~\cite{qe_tech_rep1} we represented D-sequents in the following form $(F,q,X') \rightarrow X''$. In terms of the current  paper, such a D-sequent says
that the variables of $X'$ are redundant in \prob{X}{\cof{F}{q}} and the variables of $X''$ are redundant 
in $\prob{X}{\cof{F}{q} \setminus \Sup{(\cof{F}{q})}{X'}}$. The flaw of this definition is that redundancy of variables of $X''$ is predicated
on that of variables of some other set $X'$. To solve this problem, in ~\cite{qe_tech_rep3}, we changed the definition of a D-sequent
representing it in the form $(\prob{X}{F},q) \rightarrow Z$. Such a D-sequent says that the variables of $Z$ are redundant in \prob{X}{\cof{F}{q}}.
The drawback of such definition is that it ignores the fact that variables redundant in \prob{X}{\cof{F}{q}} may not be redundant
in \prob{X}{\cof{F}{s}} where $q \subseteq s$ (see footnote~\ref{fnote:var_red}). Definition~\ref{def:d_sequent} of this paper takes care of both
problems above. First, redundancy of variables of $Z$ is not predicated on that  of some other set of variables. 
Second, by forbidding to make assignments to scope variables $W$
we guarantee that variables redundant in \prob{X}{\cof{F}{q}} are   redundant in \prob{X}{\cof{F}{s}} where $q \subseteq s$.

}.


%
%
\begin{example}
Consider an \ecnf{} formula \prob{X}{F} where $F=C_1 \wedge C_2$, 
$C_1=x \vee y_1$ and $C_2=\overline{x} \vee y_2$ and  $X =\{x\}$.
Let \pnt{q}=\s{(y_1=1)}. Then \cof{F}{q} = $C_2$ because $C_1$ is satisfied.
Notice that  $x$ is monotone and so locally redundant in \cof{F}{q} (Proposition~\ref{prop:bl_var_red}).
Hence, the D-sequent \DDs{X}{F}{q}{\s{x}}{x} holds.
\end{example}

According to Definition~\ref{def:d_sequent}, a D-sequent holds with respect to a particular \ecnf{} formula \prob{X}{F}.
Proposition~\ref{prop:form_replacement} shows
that this D-sequent also holds after adding to $F$ resolvent clauses.

%
%
\begin{proposition}
\label{prop:form_replacement}
Let \prob{X}{F} be an \ecnf{} formula. Let $H = F \wedge G$ where $F \Rightarrow G$.
Let \pnt{q} be an assignment to \V{F}.
Then if \Dss{X}{F}{q}{W}{Z} holds, \Dss{X}{H}{q}{W}{Z} does too.
\end{proposition}
The proposition below shows that it is safe to increase the scope of a D-sequent.
%
%
\begin{proposition}
\label{prop:increase_scope}
Let D-sequent \Dss{X}{F}{q}{W}{Z} hold. Let $W'$ be a superset of $W$ where  $W' \cap \Va{q} = \emptyset$.
Then \Dss{X}{F}{q}{W'}{Z} holds as well. 
\end{proposition}
%
%
\subsection{Join Operation for D-sequents}
\label{subsec:res_d_seqs}
In this subsection, we introduce the operation of joining D-sequents.
The join operation produces a new D-sequent from two D-sequents derived earlier. 

%
%
\begin{definition}
\label{def:res_part_assgns}
Let \pnt{q'} and \pnt{q''} be  assignments in which exactly  one variable $v \in \Va{q'} \cap \Va{q''}$ is assigned different values.
The assignment \pnt{q} consisting of all  the assignments of \pnt{q'} and \pnt{q''} but those to $v$ is called the
 \tb{resolvent} of \pnt{q'},\pnt{q''} on $v$.
Assignments \pnt{q'},\pnt{q''} are called \tb{resolvable}  on $v$.
\end{definition}
%
%
\begin{proposition}
\label{prop:join_rule}
Let \prob{X}{F} be an \ecnf{} formula. Let D-sequents   $(\prob{X}{F},\pnt{q'},W')$ $\rightarrow Z$  and \Dss{X}{F}{q''}{W''}{Z}
hold and $(\Va{q'} \cap W'')=(\Va{q''} \cap W')=\emptyset$. Let \pnt{q'}, \pnt{q''}
be resolvable on $v \in \V{F}$ and \pnt{q} be the resolvent of \pnt{q'} and \pnt{q''}.
Then, the D-sequent  \Dss{X}{F}{q}{W' \cup W''}{Z}  holds  too.
\end{proposition}
%
%
\begin{definition}
\label{def:join_rule}
We will say that the D-sequent \Dss{X}{F}{q}{W' \cup W''}{Z} \ of Proposition~\ref{prop:join_rule} is produced 
by \tb{joining D-sequents} \Dss{X}{F}{q'}{W'}{Z} and \Dss{X}{F}{q''}{W''}{Z}
at $v$. 
\end{definition}
%
%
\subsection{Composable D-sequents}
\label{subsec:compos_dseqs}
In general, the fact that D-sequents \DDs{X}{F}{q}{W}{v'} and  $(\prob{X}{F},\pnt{q},W)$ $\rightarrow \s{v''}$ hold does not imply
that  \DDs{X}{F}{q}{W}{v',v''} does too. The reason is that derivation of D-sequent  \DDs{X}{F}{q}{W}{v',v''}
may involve recursive reasoning where \s{v'}-clauses  are used to prove redundancy of variable $v''$ and vice versa.
Proposition~\ref{prop:merge_dseqs} below shows how to avoid recursive reasoning.

%
%
\begin{definition}
\label{def:consist_assignments}
Let \pnt{q'} and \pnt{q''} be assignments to a set of variables $Z$.
We will say that \pnt{q'} and \pnt{q''} are  \tb{compatible} if 
every variable of $\Va{q'} \cap \Va{q''}$ is assigned the same value
in \pnt{q'} and \pnt{q''}.
\end{definition}

%
%
\begin{proposition}
\label{prop:merge_dseqs}
Let \pnt{s} and \pnt{q} be assignments to variables of $F$ where $\pnt{s} \subseteq \pnt{q}$.
Let D-sequents \Dss{X}{F}{s}{W}{Z} and \DDs{X}{F \setminus F^Z}{q}{\s{v}}{v} hold where $\Va{q} \cap Z$
= $\Va{q} \cap W = \emptyset$.
Then D-sequent $(\prob{X}{F},\pnt{q},{W \cup \s{v}})$ $\rightarrow {Z \cup \s{v}}$ holds.
\end{proposition}

%
%
\begin{definition}
\label{def:composable}
Let $S'$ and $S''$ be D-sequents \Dss{X}{F}{q'}{W}{Z} and \\ \DDs{X}{F}{q''}{\s{v}}{v} respectively
where \pnt{q'} and \pnt{q''} are compatible assignments to \V{F} and $v \not\in \Va{q'}$,$\Va{q''} \cap Z = \emptyset$,
$\Va{q'} \cap W = \emptyset$.
We will call $S'$ and $S''$  \tb{composable} if
D-sequent $S$ equal to $(\prob{X}{F},\pnt{q},{W \cup \s{v}})$ $\rightarrow {Z \cup \s{v}}$
holds where $\pnt{q} = \pnt{q'} \cup \pnt{q''}$.
From Proposition~\ref{prop:merge_dseqs} it follows that   if
D-sequent \DDs{X}{F \setminus F^Z}{q}{\s{v}}{v} holds, then $S',S''$ are composable.
\end{definition}

\section{Description of \Di}
\label{sec:alg_descr}
In this section, we describe  a QE algorithm called \Di (Derivation of D-Sequents). \Di derives  D-sequents 
 \DDs{X}{F}{s}{W}{x} stating the redundancy of one variable of $X$. 
We will call D-sequent \DDs{X}{F}{s}{W}{x} \tb{active} in the branch specified by
assignment \pnt{q} if $\pnt{s} \subseteq \pnt{q}$ i.e. if this D-sequent provides a proof of redundancy of $x$ in subspace \pnt{q}.
From now on, we will use a short notation of  D-sequents writing \Dds{s}{\s{x}} instead of \DDs{X}{F}{s}{W}{x}.  We will assume 
that the parameter  \prob{X}{F} missing in \Dds{s}{\s{x}} is  the \ti{current} \ecnf{} formula (with all resolvent clauses added to $F$ so far).
We will also assume that the missing parameter $W$ is the set of variables that are currently redundant. 
One can omit \prob{X}{F} from D-sequents because from  Proposition~\ref{prop:form_replacement} it follows that once  D-sequent \DDs{X}{F}{s}{W}{x}
 is derived it holds after adding any set of resolvent clauses to $F$.  The scope parameter $W$ can be dropped because
Proposition~\ref{prop:increase_scope} entails that it is safe to increase the scope of a D-sequent. So one can just assume that
all the D-sequents that are currently active have the same scope equal to  the current set of redundant variables.

%
%
%
\setlength{\intextsep}{5pt}
\setlength{\textfloatsep}{10pt}
\begin{figure}
\small
\vspace{10pt}
\begin{tabbing}
aaaa\=bb\= c\= dddddddddddd\= \kill
// $\Phi$ denotes  \prob{X}{F},  \pnt{q} is an assignment to \V{F} \\
// \DS~denotes a set of active D-sequents \\
\Di\!\!($\Phi$,\pnt{q},\DS)\{\\
\tb{\scriptsize{1}}\> $(\DS,ans,C) \leftarrow atomic\_D\_seqs(\Phi,\pnt{q},\DS)$; \\
\tb{\scriptsize{2}}\> if (\ti{ans} = \ti{sat}) return($\Phi,\DS,\mi{sat}$);\\
\tb{\scriptsize{3}}\>if (\ti{ans} = \ti{unsat}) return($\Phi,\DS,\mi{unsat},C$);\\
\tb{\scriptsize{4}}\> $v := \mi{pick\_variable}(F,\pnt{q},\DS)$; \\
\tb{\scriptsize{5}}\> $(\Phi,\DS,\mi{ans}_0,C_0) \leftarrow$\DI($\Phi$,$\pnt{q}\cup \s{(v=0)}$,\DS);\\
\tb{\scriptsize{6}}\> $(\Sup{\DS}{sym},\Sup{\DS}{asym}) \leftarrow \mi{split}(F,\DS,v)$;\\
\tb{\scriptsize{7}}\> if ($\Sup{\DS}{asym} = \emptyset$)  return($\Phi,\DS,\mi{ans}_0,C_0$);\\
\tb{\scriptsize{8}}\> $\DS := \DS \setminus \Sup{\DS}{asym};$ \\
\tb{\scriptsize{9}}\> $(\Phi,\DS,\mi{ans}_1,C_1) \leftarrow$\DI($\Phi$,$\pnt{q} \cup \s{(v=1)}$,\DS);\\
\tb{\scriptsize{10}}\> if (($\mi{ans}_0= \mi{unsat}$) and ($\mi{ans}_1= \mi{unsat}$))\{\\
\tb{\scriptsize{11}}\Tt  $C := resolve\_clauses(C_0,C_1,v)$; \\
\tb{\scriptsize{12}}\Tt  $F := F \wedge C$; \\
\tb{\scriptsize{13}}\Tt$\DS :=\mi{process\_unsat\_clause}(\Phi,C,\DS)$;\\
\tb{\scriptsize{14}}\Tt  return($\Phi,\DS,\mi{unsat},C$);\} \\
\tb{\scriptsize{15}}\> $\DS := \mi{merge}(\Phi,\pnt{q},v,\Sup{\DS}{asym},\DS)$;\\
\tb{\scriptsize{16}}\> return($\Phi,\DS,\mi{sat}$);\} \\
\end{tabbing} 
\vspace{-20pt}
\caption{\Di procedure}
\label{fig:high_level_descr}
\end{figure}

%

A  description of \Di is given in Figure~\ref{fig:high_level_descr}.
\Di accepts an \ecnf{} formula \prob{X}{F} (denoted as $\Phi$), an assignment \pnt{q} to \V{F} and a set \DS~of  active D-sequents
stating redundancy of \ti{some} variables of  $X \setminus \Va{q}$ in \prob{X}{\cof{F}{q}}. \Di returns a modified formula
\prob{X}{F} (where resolvent clauses have been added to $F$) and a  set \DS~of active D-sequents stating redundancy of \ti{every}
variable of $X \setminus \Va{q}$ in \prob{X}{\cof{F}{q}}. \Di also returns the answer \ti{sat} if \cof{F}{q}
is satisfiable. If \cof{F}{q} is unsatisfiable, \Di returns the answer \ti{unsat} and  a clause of $F$ falsified by \pnt{q}.
To build a CNF formula  equivalent to $\Phi$, one needs to call \Di with $\pnt{q} = \emptyset$, $\DS = \emptyset$ and discard the  $X$-clauses
of the CNF formula  $F$  returned by \Di\!\!.

\subsection{The Big Picture}
First, \Di looks for variables whose redundancy is trivial to prove 
(lines 1-3). If some  variables of $X \setminus \Va{q}$ are not proved redundant yet, \Di picks a branching variable $v$ (line 4).
Then it extends  \pnt{q} by assignment  $(v=0)$ and recursively calls itself (line 5) starting the left branch of $v$.
Once the left branch is finished,
\Di extends \pnt{q} by $(v=1)$  and explores the right branch (line 9). The results of the left and right branches are
then merged (lines 10-16).

\Di terminates when, for every variable $x$ of $X \setminus \Va{q}$, it derives a D-sequent $\pnt{s} \rightarrow \s{x}$ where $\pnt{s} \subseteq \pnt{q}$.
As we show in the appendix (see Lemma~\ref{lemma:compos_dseqs}) D-sequents derived by \Di are composable.
Thus derivation of D-sequents for individual variables also  means  that a D-sequent \Dds{\pnt{s^*}}{(X \setminus \Va{q})} holds 
where $\pnt{s^*} \subseteq \pnt{q}$. So, \Di terminates when the QE problem is solved for $\Phi$ in  subspace \pnt{q}. 
The composability of D-sequents is achieved by \Di by guaranteeing that
\begin{itemize}
\item for every path of the search tree leading to a leaf, variables are proved redundant in a particular order
(but for different paths the order may be different);
\item all the \s{v}-clauses are marked as redundant and ignored as long as variable $v$ stays redundant.
\end{itemize}
So there is no path leading to a leaf of the search tree
on which recursive reasoning is employed where \s{v'}-clauses are used to prove redundancy of $v''$ and vice versa.

%
%
\subsection{Building Atomic D-sequents}
\label{subsec:atomic_d_sequents}
%
%
%
\setlength{\intextsep}{2pt}
\setlength{\textfloatsep}{4pt}
\begin{figure}
\small
\begin{tabbing}
aaa\=bb\= c\= ddddddd\= \kill
$\mi{atomic\_D\_seqs}(\Phi,\pnt{q},\DS)$\{\\
\tb{\scriptsize{1}} \> if ($\exists$ clause $C \in F$ falsif. by \pnt{q})\{ \\ 
\tb{\scriptsize{2}}\Tt$\DS\!:=\!\mi{process\_unsat\_clause}(\Phi,C,\DS)$;\\
\tb{\scriptsize{3}} \Tt   return($\DS,\mi{unsat},C$);\}\\
\tb{\scriptsize{4}}\>\DS:=\ti{new\_redund\_vars}($\Phi$,\pnt{q},\DS);\\
\tb{\scriptsize{5}}\> if ($\mi{all\_unassgn\_vars\_redund}(\Phi,\pnt{q},\DS)$)  return($\DS,\mi{sat}$);\\
\tb{\scriptsize{6}}\> return($\DS,\mi{unknown}$)\};\\
\end{tabbing}
\vspace{-25pt} 
\caption{\ti{atomic\_D\_seqs} procedure}
\vspace{5pt}
\label{fig:atomic_d_seqs}
\end{figure}

%
Procedure \ti{atomic\_D\_seqs} is called by \Di to compute D-sequents for trivial cases of variable redundancy listed 
in Section~\ref{sec:trivial_cases}. We refer to such D-sequents as \tb{atomic}. Procedure \ti{atomic\_D\_seqs}
returns an updated set of active D-sequents \DS~and answer \ti{sat}, \ti{unsat}, or \ti{unknown} depending on whether  $F$ is satisfiable, unsatisfiable
or its satisfiability is not known yet. If $F$ is unsatisfiable, \ti{atomic\_D\_seqs} also returns
 a clause $C$  of $F$ falsified by the current assignment \pnt{q}. 

Lines 1-3 of  Figure~\ref{fig:atomic_d_seqs} show what is done
when $F$ contains a clause $C$ falsified by \pnt{q}. In this case, 
every unassigned variable of $F$ becomes redundant (Proposition~\ref{prop:unsat_clause}). So, 
for every variable of $x \in X \setminus \Va{q}$ for which \DS~does not contain a D-sequent yet,
procedure \ti{process\_unsat\_clause} generates  D-sequent \Dds{s}{\s{x}} and adds it to \DS. Here
\pnt{s} is the shortest assignment falsifying $C$. 
Once \DS~contains a D-sequent for every variable of $X \setminus \Va{q}$,
 \ti{atomic\_D\_seqs} terminates returning the answer \ti{unsat}, set \DS~and clause $C$.

If no  clause of $F$ is falsified by \pnt{q}, for every variable $x$ of $X \setminus \Va{q}$ 
that does not have a D-sequent in \DS~and that is blocked, 
a D-sequent is built  as explained below. This D-sequent is then added to \DS~(line 4). If every variable of $X \setminus \Va{q}$
 has a D-sequent in \DS, then \cof{F}{q} is satisfiable. (If \cof{F}{q} is \ti{unsatisfiable}, 
variables of $X \setminus \Va{q}$ can be made redundant \ti{only} by adding a clause falsified by \pnt{q}.)
So, \ti{atomic\_D\_seqs} returns the answer \ti{sat} and set \DS~(line 5).

Given a blocked variable $x \in X \setminus \Va{q}$ of \cof{F}{q}, a D-sequent \Dds{s}{\s{x}} is built as follows.
The fact that $x$ is blocked in \cof{F}{q} means that  for any pair of clauses $C'$,$C''$ resolvable on $x$, $C'$ or $C''$  is 
either satisfied by \pnt{q} or redundant (as containing a variable proved redundant in \prob{X}{\cof{F}{q}} earlier). Assume for the sake
of clarity that it is always clause $C'$.
 The assignment  \pnt{s} 
is a subset of \pnt{q} guaranteeing that every clause $C'$ remains satisfied by \pnt{s} or redundant in \prob{X}{\cof{F}{s}} and
so $x$ remains blocked in \cof{F}{s}. If $C'$  is satisfied
by \pnt{q}, then \pnt{s} contains a single-variable assignment of \pnt{q} satisfying $C'$. If $C'$ is not satisfied
by \pnt{q} but contains a variable $x^*$ proved redundant earlier, \pnt{s} contains all the single-variable assignments of \pnt{s^*}
where \Dds{s^*}{\s{x^*}} is the D-sequent of \DS~stating redundancy of $x^*$.

%
%
\subsection{Selection of a Branching Variable}
\label{subsec:br_var_sel}

 Let  \pnt{q} be the  assignment  \Di is called with and $\Sub{X}{red}$
 be the set of variables of $X$ whose D-sequents are in the current set \DS.  Let $Y = \V{F} \setminus X$.
 \Di branches only on a subset of free (\ie unassigned) variables of $X$ and $Y$.
 Namely, a variable $x \in X \setminus \Va{q}$ is picked for branching only if $x \not\in \Sub{X}{red}$.
 A variable $y \in Y \setminus \Va{q}$ is picked for branching only if it is not detached. A variable $y$ of $Y \setminus \Va{q}$ is called
\tb{detached} in  \cof{F}{q}, if every \s{y}-clause $C$ of \cof{F}{q} that has  at least one variable of $X$ is redundant
 (because $C$ contains a variable of \Sub{X}{red}). 

 Although Boolean Constraint Propagation (BCP) is not shown explicitly
in Figure~\ref{fig:high_level_descr}, it is included into the \ti{pick\_variable} procedure as follows:
a) preference is given  to branching on variables of unit clauses of \cof{F}{q} (if any); b) if $v$ is a variable of a unit clause of $C$ of \cof{F}{q}
and $v$ is picked for branching, then the value falsifying $C$ is assigned first to cause immediate termination of this branch.
 In the description of \Di we give in  Figure~\ref{fig:high_level_descr}, the left branch always explores assignment $v=0$ but obviously $v=1$
can be explored first too.

To simplify making the branching variable $v$ redundant when merging results of the left and right branches,
\Di first assigns values to variables of $Y$ (more details are given in Subsection~\ref{subsec:branch_merging}).  This means that \ti{pick\_variable}
never selects a  variable $x \in X$ for branching, if there is a free  non-detached variable of $Y$. 
In particular, BCP does not assign values to variables of $X$ if a non-detached variable of $Y$  is still unassigned.

%
%
\subsection{Switching from Left to Right Branch}
\label{subsec:left_to_right}
\Di prunes big chunks of the search space by not branching on redundant variables of $X$.
One more powerful pruning technique of \Di
discussed in  this subsection  is to reduce  the size of right branches.

Let \Dds{s}{\s{x}} be a  D-sequent of  the set $\DS$~
computed by \Di in the left branch $v=0$ (line 5 of Figure~\ref{fig:high_level_descr}).
 Notice that if \pnt{s} has no assignment $(v\!\!=\!\!0)$, variable $x$ remains redundant
in \prob{X}{\cof{F}{q_1}}  where $\pnt{q_1} = \pnt{q} \cup \s{(v=1)}$. This is because  \Dds{s}{\s{x}} is still 
active in subspace \pnt{q_1}. 
\Di splits the set $\DS$ into subsets \Sup{\DS}{sym} and \Sup{\DS}{asym} of  D-sequents symmetric and asymmetric
with respect to variable $v$ (line 6). We call a D-sequent \Dds{s}{\s{x}} \ti{symmetric} with respect to $v$,
 if \pnt{s} does not contain an assignment to $v$ and \ti{asymmetric} otherwise.

Denote by \Sup{X}{sym} and \Sup{X}{asym} the variables of $\Sub{X}{red} \setminus \Va{q}$
whose redundancy is stated by D-sequents of  \Sup{\DS}{sym} and \Sup{\DS}{asym} respectively.
Before exploring the right branch (line 9), the variables of \Sup{X}{asym}
become non-redundant again.   Every clause $C$ of \cof{F}{q} with a variable of \Sup{X}{asym} is  unmarked as currently non-redundant unless
$\V{C} \cap \Sup{X}{sym} \neq \emptyset$.

Reducing the set of free variables of the right branch to \Sup{X}{asym} allows to prune big parts of the search space.
In particular, if \Sup{X}{asym} is empty there is no need to explore the right branch. In this case, \Di just returns
the results of the left branch (line 7). Pruning the right branch when \Sup{X}{asym} is empty is similar to non-chronological
backtracking well known in SAT-solving~\cite{grasp}.

%
%
\subsection{Branch Merging}
\label{subsec:branch_merging}
Let  $\pnt{q_0}=\pnt{q} \cup \s{(v=0)}$ and $\pnt{q_1}=\pnt{q} \cup \s{(v=1)}$.
The goal of branch merging is to extend the redundancy of all unassigned variables of $X$ proved in \prob{X}{\cof{F}{q_0}} and \prob{X}{\cof{F}{q_1}} 
to formula \prob{X}{\cof{F}{q}}.
If both \cof{F}{q_0} and \cof{F}{q_1} turned out to be unsatisfiable, this
is done as described in lines 11-14 of Figure~\ref{fig:high_level_descr}. 
In this case, the unsatisfied clauses $C_0$ and $C_1$ of \cof{F}{q_0} and \cof{F}{q_1} returned in the left and right branches respectively
are resolved on  $v$.  The resolvent $C$ is added to $F$.  
Since $F$ contains a clause $C$ that is falsified
by \pnt{q}, for every variable $x \in X \setminus \Va{q}$ whose D-sequent is not in \DS,
\Di derives an atomic D-sequent and adds it  to \DS.  This is performed by procedure
\ti{process\_unsat\_clause}  described in Subsection~\ref{subsec:atomic_d_sequents}.
If, say,  $v\!\not\in\!\V{C_1}$, then \ti{resolve\_clauses} (line 11) returns $C_1$ itself since $C_1$ is falsified by \pnt{q}
and no new clause is added to $F$.

%
%
%
\setlength{\intextsep}{4pt}
\setlength{\textfloatsep}{4pt}
\begin{figure}
\small
\vspace{10pt}
\begin{tabbing}
aaa\=bb\= cc\= ddddddd\= \kill
$\mi{merge}(\Phi,\pnt{q},v,\Sup{\DS}{asym},\DS)$\{\\
\tb{\scriptsize{1}}\> $\DS := \mi{join\_D\_seqs}(v,\Sup{\DS}{asym},\DS)$; \\
\tb{\scriptsize{2}}\> if ($v \in X$)  $\DS := \DS \cup \{\mi{atomic\_D\_seq\_for\_v}(F,\pnt{q},v,\DS)\}$;\\  
\tb{\scriptsize{3}}\> return($\DS$);\} \\
\end{tabbing} 
\vspace{-30pt}
\caption{\ti{merge} procedure}
\vspace{5pt}
\label{fig:merge}
\end{figure}

%

If at least one branch returns answer \ti{sat}, then \Di calls procedure \ti{merge} described in Figure~\ref{fig:merge}. First,
\ti{merge} takes care of the variables of  \Sup{X}{asym} 
(see Subsection~\ref{subsec:left_to_right}).
Note that redundancy of  variables of \Sup{X}{asym} is already proved in both branches. 
If a D-sequent of a variable from \Sup{X}{asym} returned in the \ti{right} branch is asymmetric in $v$,
then \ti{join\_D\_seqs} (line 1)  replaces it with a D-sequent symmetric in $v$  as follows.
Let $x \in \Sup{X}{asym}$ and  $S_0$ and $S_1$ be the D-sequents stating the redundancy of $x$ derived in the left and right branches
respectively. Procedure \ti{join\_D\_seqs} joins $S_0$ and $S_1$ at $v$ producing a new D-sequent $S$. The latter also states the redundancy of  $x$ 
but is symmetric in $v$. D-sequent $S_1$ is replaced in \DS~ with $S$.

Let us consider the case\footnote{The description of  this case given in~\cite{qe_tech_rep3} says that if $S_1$ is symmetric in $v$,
it remains in \DS~untouched. It is an error because, as we mentioned above,  the set of D-sequents produced  for subspace \pnt{q}
may turn out to be uncomposable.
} where  $S_1$ is symmetric in $v$. If \cof{F}{q_0}
 was unsatisfiable, then $S_1$ remains in \DS~ untouched. Otherwise,
\ti{join\_D\_seqs} does the following.  Let $S_1$ be equal to \Ds{s}{x}. First, 
the right branch assignment $v=1$ is added to \pnt{s}, which makes $S_1$ asymmetric in $v$.
 Then $S_1$ is joined with $S_0$ at $v$ to produce
a new D-sequent $S$ that is symmetric in $v$. $S$ replaces $S_1$ in \DS. The reason one cannot simply keep   
$S_1$  in \DS~untouched is as follows. As we mentioned above, the composability of D-sequents built by \Di~is based on the assumption that 
for every path of the search tree, variables are proved redundant in a particular order. Using D-sequent $S_1$ in subspace \pnt{q} would
violate this assumption and so would break the composability of D-sequents.

Finally, if the branching variable $v$ is in $X$, \Di derives a  D-sequent stating the redundancy of $v$. Notice that 
$v$ is not currently redundant in \prob{X}{\cof{F}{q}} because \Di does not branch on redundant variables.
 As we mentioned in Subsection~\ref{subsec:br_var_sel}, the variables of $Y = \V{F} \setminus X$ are
assigned in \Di before those of $X$. This means that before $v$ was selected for branching, all free non-detached variables of $Y$ had been
assigned. Besides, every variable  of $X \setminus \Va{q}$ but $v$ has just been proved redundant in \prob{X}{\cof{F}{q}}.
 So, \cof{F}{q} may have only two types of non-redundant clauses: a) clauses having 
only detached variables of $Y$; b) unit clauses depending on $v$.  Moreover, these unit clauses cannot contain literals
of both polarities of $v$ because \ti{merge} is called only when either branch $v=0$ or $v=1$  is satisfied. Therefore, 
$v$ is monotone. So, \ti{merge} builds
an atomic D-sequent $S$ stating the redundancy of $v$  as described in Subsection~\ref{subsec:atomic_d_sequents}
and adds it to \DS~(line 2). Then \ti{merge} terminates returning \DS.

%
\subsection{Correctness of \Di}
Let \Di be  called on formula $\Phi=\prob{X}{F}$ with $\pnt{q}=\emptyset$ and $\DS=\emptyset$.
Informally, \Di is correct because a) the atomic D-sequents built by \Di are correct; b) joining D-sequents 
produces a correct D-sequent; c) every clause  added to formula $F$ is produced by resolution and so is 
 implied by $F$;d) by the time \Di backtracks to the root of the search tree, for every variable $x \in X$, 
D-sequent $\emptyset \rightarrow \s{x}$ is derived;
e) the D-sequents derived by \Di are composable, which  implies that the D-sequent $\emptyset \rightarrow X$ 
holds for the  formula \prob{X}{F}  returned by \Di\!\!.

\begin{proposition}
\label{prop:correctness}
\Di is sound and complete.
\end{proposition}
\subsection{A Run of \Di on a Simple Formula}
\label{subsec:example}

%
%
%

\begin{wrapfigure}{l}{1.3in}
 \begin{center}
    \includegraphics[width=1.1in]{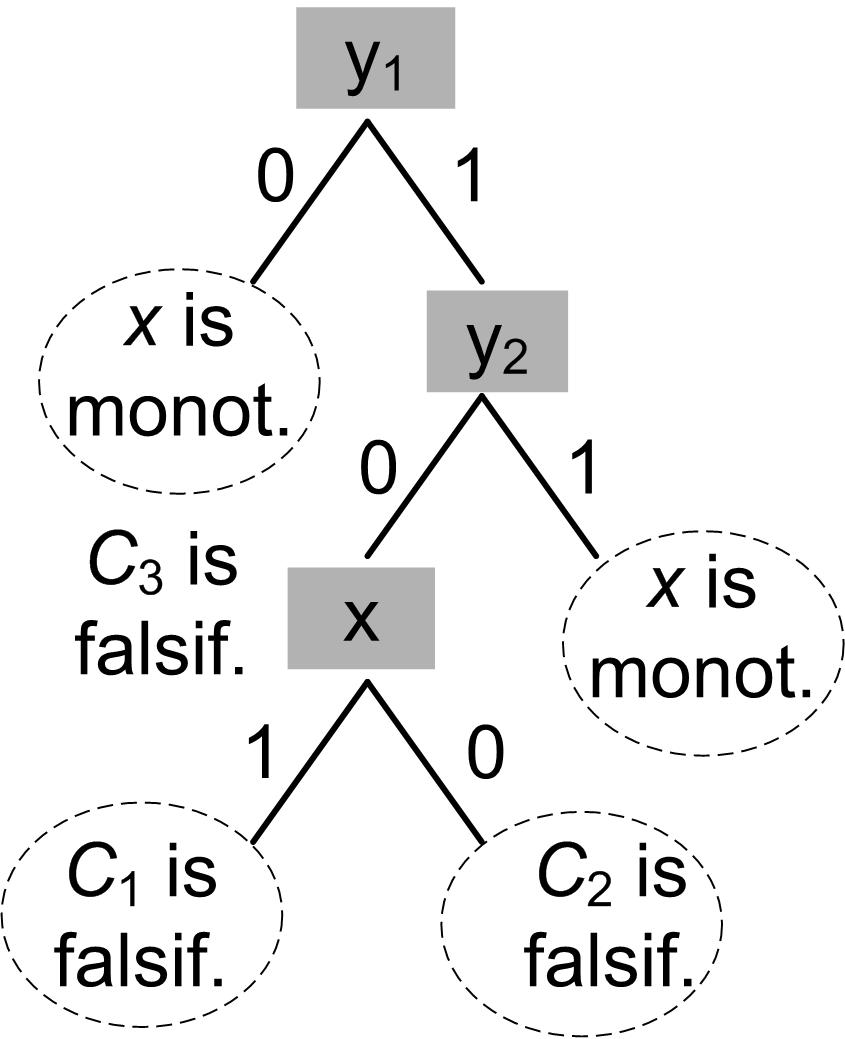}
  \end{center}
\caption{Search tree built by \Di}
\label{fig:search_tree}
\end{wrapfigure}

\begin{wrapfigure}{l}{1.9in}
 \begin{center}
    \includegraphics[height=40mm]{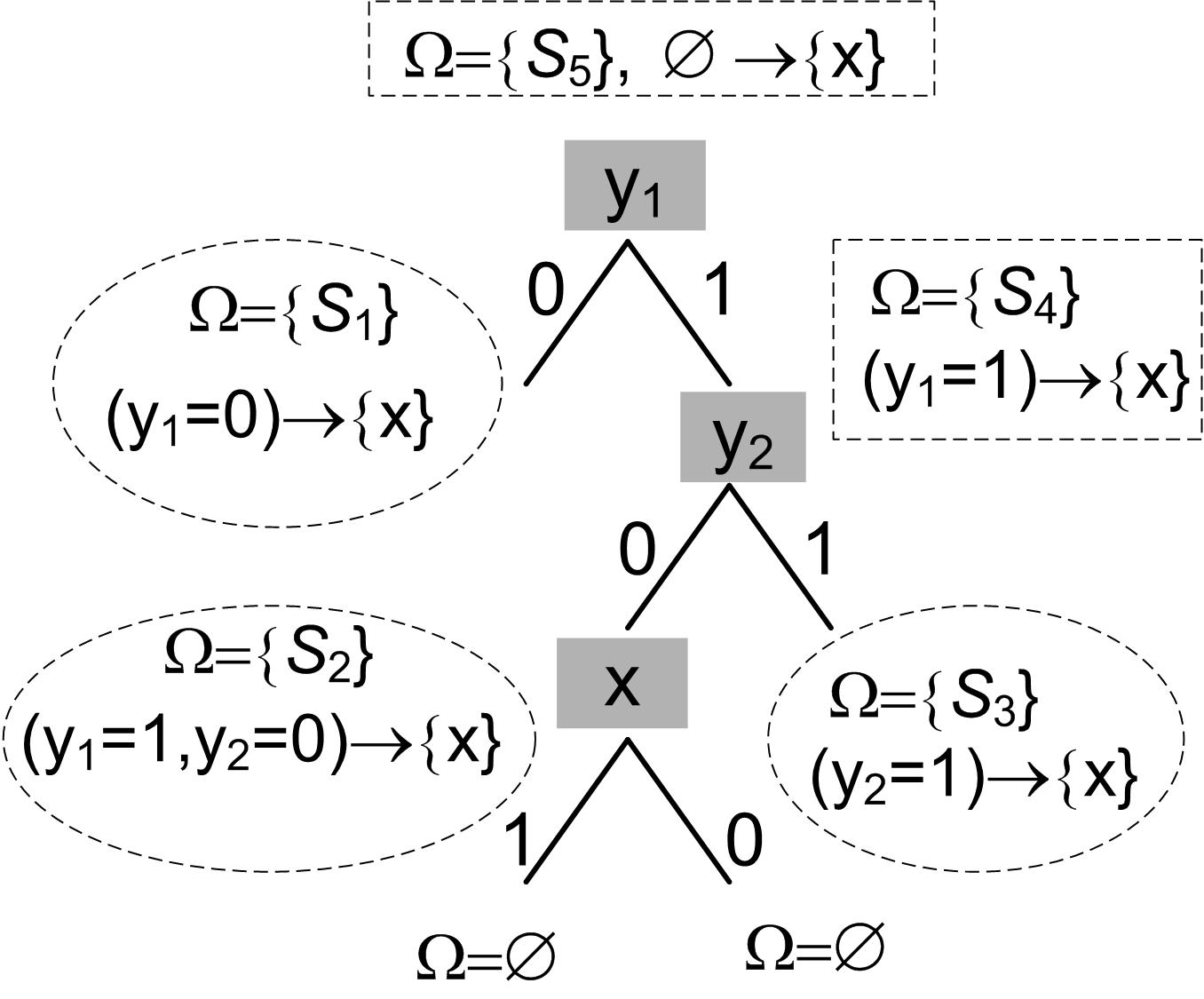}
  \end{center}
\caption{Derivation of D-sequents}
\label{fig:der_d_seqs}
\end{wrapfigure}

Let \prob{X}{F} be an \ecnf{} formula where  $F=C_1 \wedge C_2$,  $C_1 = \overline{y}_1 \vee \overline{x}$, $C_2 = y_2 \vee x$
and $X=\{x\}$. 
To identify a particular \Di call we will use the corresponding assignment \pnt{q}. For example,
\Ac{\DI}{y_1=1,y_2=0} means that 
the assignments $y_1=1$ and  $y_2=0$ were made at recursion depths 0 and 1 respectively. So the
current recursion depth is 2. 
Originally, assignment \pnt{q} is empty so the initial call is \Ac{\DI}{\emptyset}.
The work of  \Di is shown in Figures~\ref{fig:search_tree},~\ref{fig:der_d_seqs} used below to illustrate
various aspects of \Di\!\!.

\setlength{\intextsep}{4pt}

\ti{Branching variables.} Figure~\ref{fig:search_tree} shows  a search tree built by \Di\!\!.  Recall that 
\Di branches on
variables of $\V{F}\setminus X=\{y_1,y_2\}$  before those of $X$ (see Subsection~\ref{subsec:br_var_sel}).

\ti{Leaves.} The search tree of Figure~\ref{fig:search_tree} has
four leaf nodes shown in dotted ovals. 
 In each leaf node, variable  $x$ is either assigned or proved redundant.
For example, $x$  is proved redundant by \Ac{\DI}{y_1=0} and  assigned by  \Ac{\DI}{y_1=1,y_2=0,x=1}.

\ti{Generation of new clauses.} \Ac{\DI}{y_1=1,y_2=0} generates a new clause  after branching on  $x$.
\Ac{\DI}{y_1=1,y_2=0,x=1} returns $C_1$ as a clause of $F$ that is empty in \Ac{F}{y_1=1,y_2=0,x=1}. Similarly, 
\Ac{\DI}{y_1=1,y_2=0,x=0} returns $C_2$ because it  is empty in \Ac{F}{y_1=1,y_2=0,x=0}. As described in Subsection~\ref{subsec:branch_merging},
in this case, \Di resolves clauses $C_1$ and $C_2$ on the branching variable $x$.
The resolvent $C_3=\overline{y}_1 \vee y_2$ is added to $F$.

\setlength{\intextsep}{4pt}

\ti{Generation of atomic D-sequents.} Figure~\ref{fig:der_d_seqs} describes derivation of D-sequents for the search tree of Figure~\ref{fig:search_tree}. 
The atomic D-sequents are shown in dotted ovals. (Dotted boxes show D-sequents obtained by the join operation.)
For instance,  \Ac{\DI}{y_1=0}  generates D-sequent $S_1$  equal to  $(y_1\!=\!0)\rightarrow\s{x}$.
$S_1$ holds because \Ac{F}{y_1=0}=$y_2 \vee x$ and so $x$ is a blocked (monotone) variable of \Ac{F}{y_1=0}. The atomic D-sequent $S_2$ is derived
by \Ac{\DI}{y_1=1,y_2=0}. As we mentioned above, \Ac{\DI}{y_1=1,y_2=0} adds clause 
$C_3=\overline{y}_1 \vee y_2$ to $F$. This clause is empty in  \Ac{F}{y_1=1,y_2=0}. So D-sequent $S_2$ equal to $(y_1\!=\!1,y_2\!=\!0)\rightarrow \s{x}$
is generated where $(y_1\!=\!1,y_2\!=\!0)$ is the shortest assignment falsifying $C_3$.

\ti{Switching from left to right branch.} Let us consider switching between branches by  \Ac{\DI}{\emptyset} where
$y_1$ is picked for branching. The set of D-sequents  \Ac{\DS}{\emptyset} returned by the left branch equals \s{S_1} where
$S_1$ is equal to $(y_1=0)\rightarrow\s{x}$.
  The only clause $y_2 \vee x$ of \Ac{F}{y_1=0} is 
marked as redundant because it contains  $x$ that is currently redundant.
Before starting the right branch  $y_1=1$, \Ac{\DI}{\emptyset} splits \Ac{\DS}{\emptyset}
into subsets \Act{\DS}{sym}{\emptyset} and \Act{\DS}{asym}{\emptyset} of D-sequents respectively symmetric and asymmetric in $y_1$.
Since the only D-sequent of \Ac{\DS}{\emptyset}  depends on $y_1$, then \Act{\DS}{asym}{\emptyset}=\Ac{\DS}{\emptyset} and
\Act{\DS}{sym}{\emptyset}=$\emptyset$. \Ac{\DI}{\emptyset} removes D-sequent $S_1$ from \DS~
because $S_1$ becomes inactive if $y_1=1$. So, before  \Ac{\DI}{y_1=1} is called, variable $x$ becomes
non-redundant and clause $C_2 = y_2 \vee x$ is unmarked as currently non-redundant.

\ti{Branch merging.}
Consider how branch merging is performed by \Ac{\DI}{y_1=1}.  In the left branch $y_2=0$, the set  \Ac{\DS}{y_1=1}=\s{S_2} is computed where
$S_2$ is  $(y_1\!=\!1,y_2\!=\!0)\rightarrow \s{x}$. Since $S_2$ depends on  $y_2$, then \Act{\DS}{asym}{y_1=1}=\Ac{\DS}{y_1=1}.
In the right branch $y_2=1$, the set  \Ac{\DS}{y_1=1}=\s{S_3} is computed where
$S_3$ is  $(y_2=1)\rightarrow \s{x}$. By joining $S_2$ and $S_3$ at  $y_2$, D-sequent $S_4$  is derived that equals $(y_1=1)\rightarrow\s{x}$.
 $S_4$ states redundancy of $x$ in \Ac{F}{y_1=1}.

\ti{Termination.} When \Ac{\DI}{\emptyset} terminates,  $F= C_1 \wedge C_2 \wedge C_3$ where $C_3=\overline{y}_1 \vee y_2$
 and D-sequent $ \emptyset \rightarrow \s{x}$ is derived. By dropping $C_1,C_2$ as $X$-clauses one obtains $C_3 \equiv \prob{X}{C_1 \wedge C_2}$.


\section{Compositionality of \Di}
\label{sec:compos}
Let $F=F_1 \wedge \ldots \wedge F_k$ where
$\V{F_i} \cap \V{F_j}= \emptyset$, $i \neq j$. We will say that an algorithm solves the QE problem
specified by  \prob{X}{F}  \tb{compositionally} if it breaks this problem down into  
$k$ independent subproblems of finding  $G_i$ equivalent to \prob{X}{F_i}.  A formula
 $G$ equivalent to \prob{X}{F} is then built  as $G_1 \wedge \ldots \wedge G_k$.

Our interest in compositional QE algorithms is motivated as follows. First, a non-compositional algorithm has poor scalability.
Second, even if the \ti{original} formula $F$ is not a conjunction of independent subformulas,
 such subformulas may appear in subspaces of the search space
during branching. 
Notice that a QE algorithm that resolves out variables one by one as in the DP procedure~\cite{dp} is compositional.
 (Clauses of $F_i$ and $F_j$, $i \neq j$  cannot be resolved with each other).
However, such an algorithm cannot take into account  subtle properties of the formula and hence may have abysmal performance.
 Suppose, for example, that $F$ does not have independent subformulas but such subformulas 
appear in subspaces $x=0$ and $x=1$ where $x \in X$. 
A compositional \ti{branching} QE algorithm can make use of this fact in contrast to  its counterpart
eliminating quantified variables \ti{globally} i.e. for all subspaces at once.

A QE algorithm based on enumeration of satisfying assignments is not compositional. The reason is that the set of assignments satisfying $F$ is
a Cartesian product of those satisfying $F_i$,$i=1,\ldots,k$. So if, for example, all $F_i$ are identical, the complexity of an enumeration based
QE algorithm is \ti{exponential} in $k$.
 A QE algorithm
based on BDDs~\cite{bdds} is compositional only for  variable orderings where
variables of $F_i$ and $F_j$, $i \neq j$ do not interleave. 

Now we show the compositionality of \Di\!\!.   By a \ti{decision branching variable} mentioned in the proposition below,
we mean that this variable was not present in a unit clause of the current formula when it was selected for branching.

%
%
\begin{proposition}[compositionality of DDS]
\label{prop:compos}
Let $T$ be the search tree built by \Di when solving the QE  problem
\prob{X}{F_1 \wedge \ldots \wedge F_k}  $\V{F_i} \cap \V{F_j}= \emptyset$, $i \neq j$.
Let $X_i = X \cap \V{F_i}$ and $Y_i = \V{F_i} \setminus X$.
The size of  $T$ in the number of nodes
is bounded by $|\V{F}|\cdot(\eta(X_1 \cup Y_1) + \ldots +  \eta(X_k \cup Y_k))$ where 
$\eta(X_i \cup Y_i) =  2 \cdot 3^{|X_i \cup Y_i|} \cdot (|X_i| + 1), i=1,\ldots,k$
no matter how decision branching variables are chosen. 
\end{proposition}

Proposition~\ref{prop:compos}  is proved for a slightly modified version of  \Di (see the appendix of this paper). 
Notice that the compositionality of \Di is not ideal. For example, if all subformulas $F_i$ are identical, \Di is \ti{quadratic} in $k$ 
as opposed to being linear. Informally,  \Di is compositional
because D-sequents it derives have the form \Dds{s}{x} where $\Va{s} \cup \s{x} \subseteq \V{F_i}$. The only exception are D-sequents
derived when the current assignment falsifies a clause of $F$. This exception is the reason why the compositionality of \Di is not ideal.

\section{Experimental Results}
\label{sec:experiments}
 We compared \Di with a 
QE algorithm based on enumeration of
satisfying assignments~\cite{cav11} (courtesy of Andy King).  We will refer to this QE algorithm
 as \ti{EnumSA}. We also compared \Di  with the QE algorithm of~\cite{HVC} that we will call \ti{QE-GBL}.
Given a formula \prob{X}{F}, \ti{QE-GBL} eliminates variables of $X$ \ti{globally}, one by one, as in the DP procedure. However,
when resolving out a variable $x \in X$, \ti{QE-GBL} adds a new resolvent to $F$  \ti{only if} it eliminates an
\s{x}-removable \s{x}-boundary point of $F$. Variable $x$ is redundant in \prob{x}{F} if  all \s{x}-removable \s{x}-boundary points of $F$ are eliminated.
\ti{QE-GBL} does not generate  so many redundant clauses as DP, but still  has  the  flaw of eliminating variables globally.

%
%
\setlength{\abovecaptionskip}{2pt}   
\setlength{\belowcaptionskip}{2pt}   
\setlength{\intextsep}{1pt}
\setlength{\floatsep}{1pt}
\begin{wraptable}{l}{3in}
\small
\caption{\ti{Experiments with model checking formulas. The time limit is 1min}}
\vspace{-10pt}
\scriptsize
\begin{center}
\begin{tabular}{|c|c|c|c|c|c|c|} \hline
 \Ss{model che-}  &  \multicolumn{2}{|c|}{\ti{EnumSA}} & \multicolumn{2}{|c|}{\ti{QE-GBL}}
& \multicolumn{2}{|c|}{\Di} \\ 
\cline{2-3}\cline{4-5}\cline{6-7}
  \Ss{king mode}&  \Ss{solved}   &\Ss{time}      &\Ss{solved}      & \Ss{time}     &\Ss{solved}     & \Ss{time}  \\ 
        &\Ss{(\%)}  &   \Ss{(s.)}   & \Ss{(\%)}  &    \Ss{(s.)}  &\Ss{(\%)}  & \Ss{(s.)}   \\ \hline
\Ss{forward} &  \Ss{425 (56\%)}  &  \Ss{466}   & \Ss{561 (74\%)}  & \Ss{4,865} & \Ss{664 (87\%)}  & \Ss{1,530} 
 \\ \hline
\Ss{backward}  & \Ss{97 (12\%)}   &  \Ss{143}   & \Ss{522 (68\%)} & \Ss{2,744} & \Ss{563 (74\%)}  & \Ss{554} \\ \hline
\end{tabular}
\label{tbl:model_checking}
\end{center}
\end{wraptable}

 We used \ti{QE-GBL} for two reasons. First, \Di can be viewed as a branching version of \ti{QE-GBL}. So it is interesting to check
if branching is beneficial for QE algorithms.
 Second, one can consider \ti{QE-GBL} as an algorithm similar to that  of \cite{cav09}. The latter solves $\prob{x}{F(x,Y)}$ by looking for a 
Boolean function $H(Y)$ such that $F(H(Y),Y) \equiv \prob{x}{F(x,Y)}$. 
We used \ti{QE-GBL} to get an idea about the performance of the algorithm of \cite{cav09}
since it  was not implemented as a stand-alone tool.

\setlength{\intextsep}{4pt}
\begin{wrapfigure}{l}{2.5in}
 \begin{center}
 \includegraphics[width=2.5in]{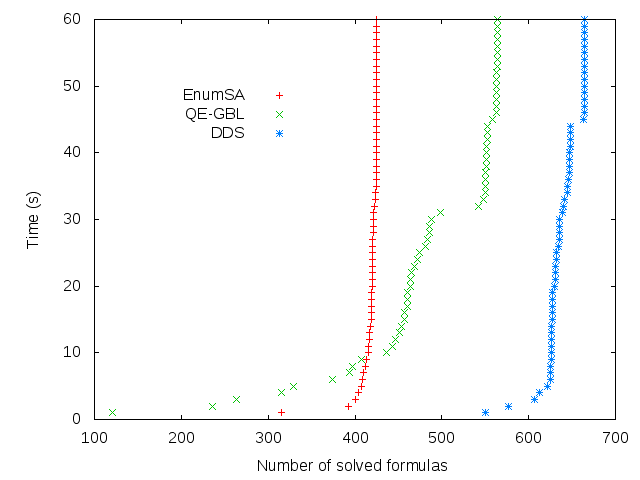}
  \end{center}
\caption{Forward model checking (1 iteration)}
\label{fig:forward_mc}
\end{wrapfigure}

Our implementation of \ti{QE-GBL} was quite efficient. In particular, we employed Picosat~\cite{picosat} for finding boundary points. 
On the other hand, in experiments, we used a very simple, proof-of-the-concept implementation of \Di\!\!. More details about this
implementation can be found in the appendix of this paper.

In the first two experiments (Table~\ref{tbl:model_checking}), we used the 758 model checking benchmarks 
of HWMCC'10 competition~\cite{hwmcc10}. In the first
experiment (the first  line of Table~\ref{tbl:model_checking}) we used  \ti{EnumSA}, \ti{QE-GBL} and \Di to compute 
the set of  states \Sub{S^1}{reach}
reachable in the first transition. In this case,  CNF formula $F$ describes the transition relation 
 and the initial state.  CNF formula $G$ equivalent to $\prob{X}{F}$
specifies  \Sub{S^1}{reach}.

In the second experiment, (the second line of Table~\ref{tbl:model_checking}) we used the same
 benchmarks to compute the set of ``bad'' states
in backward model checking. In this case, $F$ specifies the output function and
 the property in question. If $F$ evaluates to 1 for some assignment  \pnt{p} to \V{F},
 this property is broken and the state
given by the state bits of \pnt{p} is bad. Formula $G$ equivalent to $\prob{X}{F}$ specifies  the set of all bad states
(that may or may not be reachable from the initial state).

\setlength{\intextsep}{8pt}
\begin{wrapfigure}{l}{2.5in}
 \begin{center}
 \includegraphics[width=2.5in]{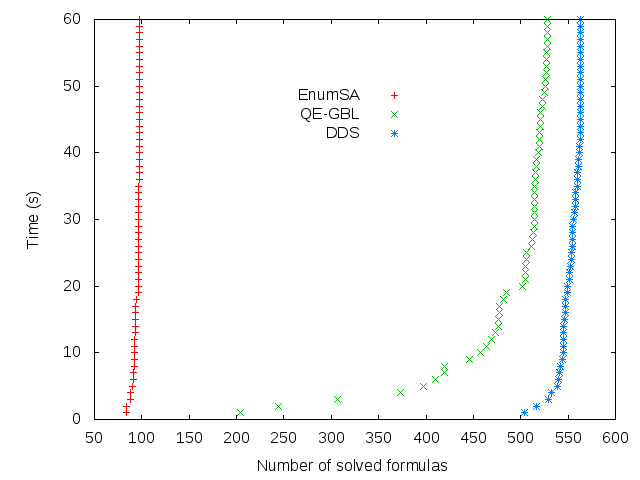}
  \end{center}
\caption{Backward model checking (1 iteration)}
\label{fig:backward_mc}
\end{wrapfigure}

Table~\ref{tbl:model_checking} shows the comparison of the three programs with respect 
to the number of formulas solved,
 percentage of this number to the total number (758) and time taken for the \ti{solved} problems.
With 1-minute time limit, \Di solved more formulas 
than \ti{EnumSA}
and \ti{QE-GBL} in forward and backward model checking.  Figures~\ref{fig:forward_mc} 
 and~\ref{fig:backward_mc} give the number of formulas of Table~\ref{tbl:model_checking}
 solved by the three programs in $t$ seconds,  $0 \leq t \leq 60$.
 These figures show the  superiority of DDS over \ti{QE-GBL} and \ti{EnumSA} on the set of formulas we used.
 The poor performance of \ti{EnumSA} on backward model checking formulas
is due to lack of constrains on next state variables. In the presence of such constraints, \ti{EnumSA} performs much better (see below).

The size of the 1,227 formulas solved by \Di peaked at 98,105 variables, the medium
size  being 2,247 variables. The largest number of non-quantified (\ie state) variables was 7,880 and
541 formulas had more than 100 state variables.  The size of  resulting formula $G$ peaked
at  32,769 clauses, 361 resulting formulas had more than 100 clauses. 
We used  Picosat~\cite{picosat} to remove redundant literals and clauses of $G$ 
with the time limit of 4 seconds. Overall, the resulting formulas  built by \Di were smaller than those of \ti{EnumSA} and \ti{QE-GBL}.
For instance, out of 1069 formulas solved by both \Di and \ti{QE-GBL}, the size of $G$ built by \Di
was  smaller  (respectively equal or larger) in 267  (respectively 798 and 4) cases.

%
%
\begin{wraptable}{l}{2.6in}
\small
\caption{\ti{Applying QE algorithms to conjunction of independent formulas. The time limit is  1 hour}}
\scriptsize
\begin{center}
\begin{tabular}{|c|c|c|c|c|c|} \hline
 \#copi- & (\#vars,           & $|Y|$ & \ti{EnumSA}   & \Di    & \Di\\ 
 es      &  \#clauses)        &       &  (s.)     &  rand (s.)  & (s.) \\ \hline
   5     &  (20,30)           & 10    &   0       &  0.01  & 0.01     \\ \hline
  10     &  (40,60)           &  20   &  10.46    &  0.01  &   0.01 \\ \hline
  15     &  (60,90)           &  30   & $>$1hour  &  0.01  &   0.01  \\ \hline
  500    &  (2000,3000)       & 1000  & $>$1hour  & 1.95   &   0.04   \\ \hline
\end{tabular}                
\label{tbl:compos_form}
\end{center}

\end{wraptable}

In the experiments above, we did not use formula 
preprocessing even though it could have been beneficial. For instance, the forward model checking formulas
had a lot of unit clauses encoding the initial state. The backward model checking formulas had many  blocked (\ie redundant) clauses~\cite{cade11}. 
The reason is that when the \ti{original} set of bad states is computed, the next state variables are not constrained yet.
However, when we compared the three programs on preprocessed formulas  we obtained similar results: \Di outperformed \ti{EnumSA} and \ti{QE-GBL}.
In particular, we generated 189 backward model checking formulas specifying bad states after a number of iterations. The idea was to get
formulas were preprocessing simplifications performing initial BCP and elimination of blocked clauses failed. 
With 1-minute time limit, \Di\!\!, \ti{QE-GBL} and \ti{EnumSA} solved 185, 163 and 149 
formulas out of 189 respectively. Notice that \ti{EnumSA} performed much better here than in the initial iteration.

The third experiment (Table \ref{tbl:compos_form}), 
clearly shows the compositionality of  \Di in comparison to \ti{EnumSA}.
 In this experiment,~both programs~computed the output~assignments produced by a combinational circuit $N$ 
composed of small \ti{identical} circuits $N_1,\ldots,N_k$ with independent sets of variables.
In this case, one needs to eliminate quantifiers from \prob{X}{F}
where $F$ = $F_1 \wedge \ldots \wedge F_k$. CNF formula $F_i$ specifies  $N_i$
 and  $\V{F_i} \setminus X$ and $\V{F_i} \cap X$  are the sets of 
output and non-output variables of $N_i$ respectively. So a CNF formula equivalent to \prob{X}{F} specifies the output assignments of  $N$.

The first column of Table~\ref{tbl:compos_form} shows  $k$ (the number of copies of $N_i$).
The next two columns give the size of CNF formula $F$ and the number of outputs in circuit $N$. 
The last three columns show the run time of \ti{EnumSA} and two versions of \Di\!\!. In the first version, the choice of
 branching variables
was random. In the second version, this choice was guided by the compositional structure of $N$. 
While \Di solved all the formulas easily, \ti{EnumSA} could not finish the formulas $F$ with $k \geq 15$ in 1 hour.
Notice that  \Di was able to
quickly solve all the formulas even with the random choice of branching variables.

\section{Background}
\label{sec:background}

The relation between a resolution proof and the process of
elimination of boundary points was discussed
in~\cite{bnd_pnts}. In terms of the present paper,
\cite{bnd_pnts} dealt only with a special kind of $Z$-boundary
points of formula $F$ where $|Z|=1$.  In the
present paper, we consider the case where $Z$ is an arbitrary
subset of the set of quantified variables $X$ of an \ecnf formula \prob{X}{F}.  This extension is crucial for describing the
semantics of D-sequents.

As far as quantifier elimination is concerned, QE algorithms and
QBF solvers can be partitioned into two categories.  (Although,
in contrast to a QE algorithm, a QBF-solver is a decision
procedure, they both employ methods of quantifier
elimination. For the lack of space, we omit references to papers
on QE algorithms that use BDDs~\cite{bdds,bdd-qe}.) The members
of the first category employ various techniques to eliminate
quantified variables of the formula one by one in some
order~\cite{beds,quantor,qubos,cav09,niclas}.  For example,
in~\cite{cav09}, quantified variables are eliminated by
interpolation. All these solvers face the same problem: there may
not exist a good \ti{single} order for variable elimination,
which, may lead to exponential growth of the size of intermediate
formulas. In Subsection~\ref{sec:compos}, we already gave an
example of this problem. Here is one more. Let \pnt{q} be an
assignment to variables of $F$.  If formula \cof{F}{q}
has unit clauses, the variables of such clauses can be eliminated
by unit resolution, \ie BCP.  In a sense, unit resolution
eliminates variables of \cof{F}{q} in a natural order.  However,
natural orders in formulas \cof{F}{q'} and \cof{F}{q''}
of different branches \pnt{q'} and \pnt{q''} may be
incompatible.

The solvers of the second category are based on enumeration of
satisfying or unsatisfying
assignments~\cite{blocking_clause,fabio,cofactoring,cav11,plaisted}.
Since such assignments are, in general, ``global'' objects, it is
hard for such solvers to follow the fine structure of the
formula, \eg  such solvers are not compositional.  In a sense, \Di tries
to take the best of both worlds. It branches and so can use
different variable orders in different branches as the solvers of
the second category. At the same time, in every branch, \Di
eliminates quantified variables individually as the solvers of
the first category, which makes it easier to follow the formula
structure.


\section{Conclusion}
\label{sec:conclusion}
We introduced Derivation of Dependency-sequents (\DI), a new
method for eliminating quantifiers from a formula \prob{X}{F} where
$F$ is a CNF formula.
The essence of \Di is to add resolvent clauses to $F$ to make the
variables of $X$ redundant. The process of making variables
redundant is described by dependency sequents (D-sequents)
specifying conditions under which variables of $X$ are
redundant. In contrast to methods based on the enumeration of
satisfying assignments, \Di is compositional.  Our experiments
with a proof-of-the-concept implementation show the promise of
\DI. Our future work will focus on studying various ways to
improve the performance of \DI, including lifting the constraint
that non-quantified variables are assigned before quantified
variables and reusing D-sequents instead of discarding them after
one join operation (as SAT-solvers reuse conflict clauses).


\section{Acknowledgment}
This work was funded  in part by NSF grant CCF-1117184 and  SRC contract 2008-TJ-1852.

\ifx \FullVersion \undefined
\else
\appendix
\setcounter{proposition}{0}
\vspace{10pt}
\noindent\tb{\large{Appendix}} \\
\vspace{10pt}

The appendix is structured as follows. In the first section, we give some details of the implementation of \Di we used in experiments.
In the following sections we provide  proofs\footnote{The proofs of this paper are similar to those of~\cite{qe_tech_rep3}.
We changed only the parts  affected by using the notion of 
scoped redundancy of variables (see Section~\ref{sec:div_conq}).
} of the propositions  listed in the paper.
We also give proofs of lemmas that are  used in the proofs of propositions.
The numbering of propositions in the appendix is the same as in  the main body of the paper.  

%
%
\section*{Some Implementation Details}
In this section, we describe some features of the implementation of \Di we used in experiments. We will refer to this implementation
as \Dii\!\!.
\begin{itemize}
%
%
\item In Figure~\ref{fig:high_level_descr}, \Di is described in terms of recursive calls. It is more convenient, to
consider \Dii as building a search tree. Let $n$ be the node of the search tree  built by \Dii at which a variable $v$ of \V{F} is assigned.
Then the depth $\mi{Depth}(n)$ of $n$ is equal to 
the  recursion depth at which variable $v$ is assigned by \Di\!\!.

%
%
\item In \Dii\!\!, we followed the common practice of
using stack for implementing branching algorithms.
 When a new node $n$ of the search tree is created, all the relevant information about $n$  is pushed on  the stack. 
When backtracking from node $n$, all the information about $n$ is popped off the stack. 
%
%
\item To make the code of \Dii easy to modify, we have not implemented optimization techniques like using watched literals to speed up BCP,
special representation of two-literal clauses and so on.
%
%
\item
In Figure~\ref{fig:high_level_descr}, a D-sequent depending on an assignment to the branching variable 
is discarded when the current \Di call terminates. On the other hand, keeping such D-sequents may be very beneficial.
The reason is that after  getting broken, a D-sequent $S$ stating redundancy of $x \in X$ may become active
 again  in a different part of the search space.
 $S$ can be used in that part of the space to avoid branching on  $x$.
This is similar to reusing conflict clauses to avoid entering the parts of the search space already proved unsatisfiable.
Nevertheless, to keep \Dii as simple as possible,  D-sequent reusing has not been implemented.
%
%
\item
In Figure~\ref{fig:high_level_descr}, if both branches are unsatisfiable, \Di adds the resolvent $C$ of clauses $C_0$ and $C_1$ falsified
in left and right branches respectively.  Recall that $C$ is falsified by the current assignment \pnt{q}.
Let $\mi{Depth}(C)$ describe the maximum recursion depth at which an assignment of \pnt{q} falsifying a literal of $C$ is made.
In \Dii\!\!, clause $C$ is not added to $F$ if another  clause $C'$ falsified by \pnt{q} can be derived later such that
 $\mi{Depth}(C') < \mi{Depth}(C)$. This is similar to the conflict clause generation procedure of a SAT-solver. In such a procedure,
all intermediate resolvents produced  in the course of generation of a conflict clause  are discarded. 

\vspace{3pt}
The condition above means that \Dii keeps
a resolvent clause $C$ only if it is empty or if in the node of the search tree located at depth $\mi{Depth}(C)$ 
\begin{itemize}
\item the left branch is currently explored or
\item the right branch is currently explored and formula $F$  was \ti{satisfiable} in the left branch.
\end{itemize}
In terms of a conflict clause generation procedure, \Dii backtracks to the closest \ti{decision} assignment of the current path of the search tree
 or to the root of the tree if the current path does not have any decision assignments.
\end{itemize}

%
%
\section*{Propositions of Section~\ref{sec:rvars_bps}: Redundant Variables, Boundary Points and Quantifier Elimination}
%
%
\begin{proposition}
A $Z$-boundary point \pnt{p} of $F$ 
is removable in \prob{X}{F},
iff one cannot turn \pnt{p} into an assignment satisfying $F$
by changing only the values of variables of $X$.
\end{proposition}
\begin{mmproof} \ti{If part}. Assume the contrary. That is \pnt{p} is not removable while no satisfying assignment can be obtained
from \pnt{p} by changing only assignments to variables of $X$. Let $Y = \V{F} \setminus X$ and
 $C$ be a clause consisting only of variables of $Y$ and falsified
by \pnt{p}. Since \pnt{p} is not removable, clause $C$ is not implied by $F$. This means that there is an assignment \pnt{s} that
falsifies $C$ and satisfies $F$. By construction, \pnt{s} and \pnt{p} have identical assignments to variables of $Y$. Thus,
\pnt{s} can be obtained from \pnt{p} by changing only values of variables of $X$. Contradiction.
\vspace{3pt}

\noindent\textit{Only if part.} Assume the contrary. That is \pnt{p} is removable but one can obtain an assignment \pnt{s} satisfying
$F$ from \pnt{p} by changing only values of variables of $X$.  Since \pnt{p} is removable, there is a clause $C$ that
is implied by $F$ and falsified by \pnt{p} and that depends only of variables of $Y$. Since \pnt{s} and \pnt{p} have
identical assignments to variables of $Y$, point \pnt{s} falsifies $C$. However, since \pnt{s} satisfies $F$, this means
that $C$ is not implied by $F$. Contradiction $\square$
\end{mmproof}
%
%
\begin{proposition}
The variables of $Z \subseteq X$ are not redundant in \prob{X}{F} iff there is an  
$X$-removable $W$-boundary point of $F$, $W \subseteq Z$.
\end{proposition}
%
%
\begin{mmproof}
Let $H$ denote $F \setminus \Sup{F}{Z}$ and $Y$ denote $\V{F} \setminus X$. Given
a point \pnt{p}, let (\pnt{x},\pnt{y}) specify the assignments of \pnt{p} to the variables of $X$ and $Y$
respectively. \\
\noindent\textit{If part.} Assume the contrary, \ie 
 there is  an $X$-removable  $W$-boundary point \pnt{p}=(\pnt{x},\pnt{y})
 of $F$ where $W \subseteq Z$ but  the variables of $Z$ are redundant and hence
 $\prob{X}{F} \equiv \prob{X}{H}$. 
 Since \pnt{p} is a boundary point, $F(\pnt{p}) = 0$. Since
\pnt{p} is removable, $\cof{(\prob{X}{F})}{y} = 0$. On the other hand, since \pnt{p} falsifies
only $W$-clauses of $F$ it satisfies $H$.  Hence $\cof{(\prob{X}{H})}{y} = 1$ and so
$\cof{(\prob{X}{F})}{y} \neq \cof{(\prob{X}{H})}{y}$. Contradiction.

\vspace{3pt}
\noindent\textit{Only if part.} 
Assume the contrary, \ie the variables
 of $Z$ are not redundant (and hence $\prob{X}{F} \not\equiv \prob{X}{H}$)
 and there does not exist an $X$-removable $W$-boundary point of $F$, $W \subseteq Z$.  
Let \pnt{y} be an assignment to $Y$ such that 
$\cof{(\prob{X}{F})}{y} \neq \cof{(\prob{X}{H})}{y}$.
One has to consider the following two cases. 
\begin{itemize}
\item   $\cof{(\prob{X}{F})}{y}=1$ and $\cof{(\prob{X}{H})}{y}=0$.
Then there exists 
an assignment \pnt{x} to $X$ such that (\pnt{x},\pnt{y}) satisfies $F$. Since every clause of $H$
is in $F$, formula $H$ is also satisfied by \pnt{p}. Contradiction.
\item $\cof{(\prob{X}{F})}{y}=0$ and $\cof{(\prob{X}{H})}{y}=1$. Then there exists an assignment \pnt{x}
to variables of $X$ such that (\pnt{x},\pnt{y}) satisfies $H$.  Since $\cof{F}{y} \equiv 0$,
point (\pnt{x},\pnt{y}) falsifies $F$. Since $H(\pnt{p})=1$ and every clause of $F$ that
is not in $H$ is an $Z$-clause, (\pnt{x},\pnt{y}) is a $W$-boundary point of $F$ where $W \subseteq Z$.
  Since $\cof{F}{y} \equiv 0$, (\pnt{x},\pnt{y}) is  an $X$-removable $W$-boundary point of $F$. Contradiction $\square$
\end{itemize}
\end{mmproof}
%
%
\section*{Propositions of Section~\ref{sec:div_conq}: Boundary Points And Divide-And-Conquer Strategy}
%
%
\begin{proposition}
Let \prob{X}{F} be an \ecnf{} formula and \pnt{q} be an  assignment to \V{F}. Let \pnt{p} be a  $Z$-boundary point 
of $F$ where $\pnt{q} \subseteq \pnt{p}$ and $Z \subseteq X$. Then if \pnt{p} is removable in \prob{X}{F} it is also removable
in \prob{X}{\cof{F}{q}}.
\end{proposition}
%
%
\begin{mmproof}
Let $Y$ denote $\V{F} \setminus X$.
Assume the contrary. That is \pnt{p} is removable in $\prob{X}{F}$ but is not removable in \prob{X}{\cof{F}{q}}. The fact that \pnt{p} is removable
in \prob{X}{F} means that there is a clause $C$ implied by $F$ and falsified by \pnt{p} that consists only of variables of $Y$.
Since \pnt{p} is not removable in \prob{X}{\cof{F}{q}}, from Proposition~\ref{prop:rem_bnd_pnt} 
it follows that an assignment \pnt{s} satisfying \cof{F}{q}
can be obtained from \pnt{p} by changing only values of variables of $X \setminus \Va{q}$.  By construction,  \pnt{p} and \pnt{s} have identical 
assignments to variables of $Y$. So \pnt{s} has to falsify $C$. On the other hand, by construction, $\pnt{q} \subseteq \pnt{s}$.  So,
the fact that \pnt{s} satisfies \cof{F}{q} implies that \pnt{s} satisfies  $F$ too. Since \pnt{s} falsifies $C$ and satisfies $F$
the former cannot be implied by the latter. Contradiction $\square$
\end{mmproof}
%
%
\begin{proposition}
Let \prob{X}{F} be a CNF formula and \pnt{q} be an assignment to variables of $F$.
Let the variables of $Z$ be  redundant in \prob{X}{\cof{F}{q}} with scope $W$ where $Z \subseteq (X \setminus \Va{q})$. Let
a variable  $v$ of $X \setminus (\Va{q} \cup Z)$ be locally redundant in $\prob{X}{\cof{F}{q} \setminus (\cof{F}{q})^Z}$.
Then the variables of $Z \cup \s{v}$ are  redundant in \prob{X}{\cof{F}{q}} with scope $W \cup \s{v}$.
\end{proposition}
%
%
\begin{mmproof}
Assume the contrary, that is the variables of $Z \cup \s{v}$ are not redundant with scope $W \cup \s{v}$. 
Then from Definition~\ref{def:scoped_red_vars} it follows that
\cof{F}{q} has a $Z'$-boundary point \pnt{p} where $Z' \subseteq Z \cup \s{v}$, $\pnt{q} \subseteq \pnt{p}$ that is 
$(W \cup \s{v})$-removable in \cof{F}{q}.
Let us consider the two possible cases:
\begin{itemize}
\item  $v \not\in Z'$ (and so $Z' \subseteq Z$).
Since \pnt{p} is $(W \cup \s{v})$-removable in \cof{F}{q}, it is also $W$-removable in \cof{F}{q}.
Hence, the variables of $Z$ are not redundant in \prob{X}{\cof{F}{q}} with scope $W$. Contradiction.
\item  $v \in Z'$ (and so $Z' \not\subseteq Z$). 
 Then \pnt{p} is a \s{v}-boundary  point of $\cof{F}{q} \setminus (\cof{F}{q})^Z$.
Indeed,  there has to be a clause  $C$ of \cof{F}{q}  falsified by \pnt{p} that contains variable $v$.
 Otherwise,  condition d) of the definition of a boundary point is broken because $v$ can be removed from $Z'$ (see Definition~\ref{def:bnd_pnt}) .

\vspace{10pt}
Let $P$ denote the set of all 
points obtained from \pnt{p} by flipping values of variables of $W \cup \s{v}$.
Let us consider the following two possibilities.
     \begin{itemize}
        \item Every point of $P$ falsifies $\cof{F}{q} \setminus (\cof{F}{q})^Z$. This means that the point \pnt{p}
           is a $\s{v}$-removable \s{v}-
               boundary point of $\cof{F}{q} \setminus (\cof{F}{q})^Z$. So $v$ is not locally redundant in  $\prob{X}{\cof{F}{q} \setminus (\cof{F}{q})^Z}$.
           Contradiction.       

\vspace{3pt}
         \item A point \pnt{d} of $P$ satisfies $\cof{F}{q} \setminus (\cof{F}{q})^Z$. Let us consider the following two cases.
            \begin{itemize}
              \item \pnt{d} satisfies \cof{F}{q}. This contradicts the fact that \pnt{p} is
               a $(W \cup \s{v})$-removable $Z'$-boundary point of \cof{F}{q}.  
             (By flipping variables of $W \cup \s{v}$    one can  obtain a point satisfying \cof{F}{q}.)

              \item \pnt{d} falsifies some clauses of \cof{F}{q}. Since \cof{F}{q} and $\cof{F}{q} \setminus (\cof{F}{q})^Z$
 are different only in $Z$-clauses, \pnt{d} is a $Z''$-boundary point of \cof{F}{q} where $Z'' \subseteq Z$.
 By construction, \pnt{p} and \pnt{d} are different only in values of variables from $W \cup \s{v}$. So, the fact that 
 \pnt{p} is a $(W \cup \s{v})$-removable $Z'$-boundary point
                            of \cof{F}{q} implies that  \pnt{d} is a $W$-removable  $Z''$-boundary point of \cof{F}{q}.
            So the variables of $Z$ are not  redundant in \cof{F}{q} with scope $W$.
            Contradiction $\square$
            \end{itemize}
     \end{itemize}
\end{itemize}
\end{mmproof}
\vspace{6pt}

\section*{Propositions of Section~\ref{sec:trivial_cases}: Two Simple Cases of Local Variable Redundancy}
%
%
\begin{lemma}
\label{lemma:twin_bnd_pnts}
Let \pnt{p} be a \s{v}-boundary point of CNF formula $G(Z)$ where $v \in Z$.
Let \pnt{p'} be obtained from \pnt{p} by flipping the value of $v$.
Then \pnt{p'} either satisfies $G$ or it is also a \s{v}-boundary point of $G$.
\end{lemma}
\begin{mmproof}
Assume the contrary, \ie \pnt{p'} falsifies a clause $C$ of $G$ that does not
have a literal of  $v$. (And so  \pnt{p'} is neither a satisfying assignment
nor a \s{v}-boundary point of $G$.) Since \pnt{p} is different from \pnt{p'}
only in the value of $v$, it also falsifies $C$. Then \pnt{p} is not
a \s{v}-boundary point of $G$. Contradiction $\square$
\end{mmproof}
%
%
\begin{proposition}
Let \prob{X}{F} be an \ecnf{} formula and \pnt{q} be an assignment to \V{F}. Let a variable $v$ of  
 $X \setminus \Va{q}$ be
blocked in \cof{F}{q}. Then $v$ is locally redundant in \prob{X}{\cof{F}{q}}.
\end{proposition}
%
%
\begin{mmproof}
Assume the contrary i.e. $v$ is not locally redundant in \prob{X}{\cof{F}{q}}.
 Then there is 
a  $v$-removable \s{v}-boundary point \pnt{p} of \cof{F}{q}. 
Note that the clauses of \cof{F}{q} falsified by \pnt{p} have the same
literal $l(v)$ of variable $v$. Let \pnt{p'} be the point obtained from \pnt{p} by flipping the value of $v$.
According to Lemma~\ref{lemma:twin_bnd_pnts}, one needs to consider only the following two cases.
   \begin{itemize}
     \item \pnt{p'} satisfies \cof{F}{q}. Since \pnt{p'} is obtained from \pnt{p} by changing only variable $v$,
      \pnt{p} is not  \s{v}-removable in \cof{F}{q}. Contradiction.
     \item \pnt{p'} falsifies only the clauses of \cof{F}{q} with literal $\overline{l(v)}$. (Point \pnt{p'}
       cannot falsify a clause with literal $l(v)$.) Then there is a pair of clauses $C$ and $C'$ of \cof{F}{q}
 falsified by \pnt{p} and \pnt{p'} respectively
       that have opposite literals only of variable $v$. Hence $v$ is not a blocked variable of \cof{F}{q}.
        Contradiction $\square$
   \end{itemize}
\end{mmproof}
%
%
\begin{proposition}
Let \prob{X}{F} be an \ecnf{} formula and \pnt{q} be an assignment to \V{F}. 
Let \cof{F}{q} have an empty clause.
Then the variables of  $X \setminus \Va{q}$ are locally redundant in \prob{X}{\cof{F}{q}}.
\end{proposition}
\begin{mmproof}
Let $X'$ denote the set $X \setminus \Va{q}$.
Assume the contrary i.e.  the variables of $X'$ are not locally redundant in \prob{X}{\cof{F}{q}}.
Then there is  an $X'$-removable $Z$-boundary point where $Z \subseteq X'$. However, the set of $Z$-boundary points of \cof{F}{q}
is empty. Indeed, on the one hand, \cof{F}{q} contains an empty clause $C$ that is falsified by any point. On the other hand,
according to Definition~\ref{def:bnd_pnt}, if \pnt{p} is a $Z$-boundary point, then $Z$ is a non-empty set that has to contain
at least one variable of every clause falsified by \pnt{p}, in particular, a variable of clause $C$ $\square$
\end{mmproof}

%
%
\section*{Propositions of Section~\ref{sec:d_sequents}: Dependency Sequents (D-sequents)}
\begin{proposition}
Let \prob{X}{F} be an \ecnf{} formula. Let $H = F \wedge G$ where $F$ implies $G$.
Let \pnt{q} be an assignment to \V{F}.
Then if \Dss{X}{F}{q}{W}{Z} holds, the D-sequent  \Dss{X}{H}{q}{W}{Z} does too.
\end{proposition}
%
%
\begin{mmproof}
Assume the contrary, \ie \Dss{X}{F}{q}{W}{Z} holds but $(\prob{X}{F},\pnt{q},W)$ $\rightarrow Z$ does not.
According to Definition~\ref{def:d_sequent}, this means  that 
variables of $Z$ are not  redundant in \prob{X}{\cof{H}{q}} with scope $W$.
That is, there is a $W$-removable $Z'$-boundary point \pnt{p}  of \cof{H}{q} where $Z' \subseteq Z$. 
The fact that the variables of $Z$ are redundant in \prob{X}{\cof{F}{q}} with scope $W$ 
means that  \pnt{p} is not a $W$-removable $Z''$-boundary point of \cof{F}{q}
where $Z'' \subseteq Z$. This can happen for the following three reasons.
\begin{itemize}
\item \pnt{p} satisfies \cof{F}{q}. Then it also satisfies \cof{H}{q} and hence cannot be a boundary point of
 \cof{H}{q}. Contradiction.
\item \pnt{p} is not  a  $Z''$-boundary point of \cof{F}{q} where $Z'' \subseteq Z$. That is \pnt{p} falsifies a
clause $C$ of \cof{F}{q} that does not contain a variable of $Z$.
 Since \cof{H}{q} also contains $C$, 
point \pnt{p} cannot be a $Z'$-boundary point of \cof{H}{q} where $Z' \subseteq Z$. Contradiction.
\item \pnt{p} is a $Z''$-boundary point of \cof{F}{q} where $Z'' \subseteq Z$ but  it is not $W$-removable in \cof{F}{q}. This means that one can 
obtain a point \pnt{s} satisfying \cof{F}{q} by flipping values of variables of $W$ in \pnt{p}.
  Since \pnt{s} also satisfies \cof{H}{q}, one has to conclude that \pnt{p} is not a $W$-removable point of \cof{H}{q}. Contradiction $\square$
\end{itemize}
\end{mmproof}

%
%
\begin{proposition}
Let D-sequent \Dss{X}{F}{q}{W}{Z} hold. Let $W'$ be a superset of $W$ where  $W' \cap \Va{q} = \emptyset$.
Then \Dss{X}{F}{q}{W'}{Z} holds as well. 
\end{proposition}
\begin{mmproof}
Assume that \Dss{X}{F}{q}{W'}{Z} does not hold. Then there is a $V$-boundary point \pnt{p} of \cof{F}{q} where $V \subseteq Z$ that
is $W'$-removable in \cof{F}{q}. Since $W \subseteq W'$, point \pnt{p} is also $W$-removable. This means that \Dss{X}{F}{q}{W}{Z} does not hold.
Contradiction $\square$
\end{mmproof}

\vspace{5pt}
%
%
\vspace{10pt}
\begin{proposition}
Let \prob{X}{F} be an \ecnf{} formula. Let D-sequents $(\prob{X}{F},\pnt{q'},W')$ $\rightarrow Z$
and  \Dss{X}{F}{q''}{W''}{Z} hold
and $(\Va{q'} \cap W'')=(\Va{q''} \cap W')=\emptyset$.
 Let \pnt{q'}, \pnt{q''}
be resolvable on $v \in \V{F}$ and \pnt{q} be the resolvent of \pnt{q'} and \pnt{q''}.
Then, the D-sequent  \Dss{X}{F}{q}{W' \cup W''}{Z}  holds  too.
\end{proposition}
\begin{mmproof}
Assume the contrary, that is D-sequent  \Dss{X}{F}{q}{W' \cup W''}{Z} does not hold and so
 the variables of $Z$ are not redundant in \prob{X}{\cof{F}{q}} with scope $W' \cup W''$.
 Then there is a $Z^*$-boundary point \pnt{p} where $Z^* \subseteq Z$ and $\pnt{q} \subseteq \pnt{p}$
 that is $(W' \cup W'')$-removable in \cof{F}{q}.
By definition of \pnt{q}, the fact that $\pnt{q} \subseteq \pnt{p}$ implies that $\pnt{q'} \subseteq \pnt{p}$ or $\pnt{q''} \subseteq \pnt{p}$.
Assume, for instance, that $\pnt{q'} \subseteq \pnt{p}$. 
The fact that  \pnt{p} is a $Z^*$-boundary point of \cof{F}{q} implies that  \pnt{p}
is also a $Z^*$-boundary point of \cof{F}{q'}. Since  \pnt{p} is $(W' \cup W'')$-removable in \cof{F}{q}
it is also $W'$-removable in \cof{F}{q'}. So 
the variables of $Z$ are not redundant in \cof{F}{q'} with scope $W'$ and D-sequent
\Dss{X}{F}{q'}{W'}{Z} does not hold.   Contradiction $\square$
\end{mmproof}

%
%
\begin{lemma}
\label{lemma:holds_in_subspace}
Let D-sequent \Dss{X}{F}{q}{W}{Z} hold and \pnt{r} be an assignment such that $\pnt{q} \subseteq \pnt{r}$
and $\Va{r} \cap W = \emptyset$. Then  D-sequent \Dss{X}{F}{r}{W}{Z} holds too.
\end{lemma}
\begin{mmproof}
Assume the contrary i.e. the variables of  $Z$ are not  redundant in \cof{F}{r} with scope $W$.
Then there is a $Z'$-boundary point \pnt{p} where $Z' \subseteq Z$ that is $W$-removable in \cof{F}{r}.
Note that \pnt{p} is also a $Z'$-boundary point of \cof{F}{q} and it is also $W$-removable in \cof{F}{q}.
This implies that the variables of $Z$ are not redundant  in \cof{F}{q} with scope $W$. Contradiction.
\end{mmproof}

%
%
\begin{proposition}
Let \pnt{s} and \pnt{q} be assignments to variables of $F$ where $\pnt{s} \subseteq \pnt{q}$.
Let D-sequents \Dss{X}{F}{s}{W}{Z} and \DDs{X}{F \setminus F^Z}{q}{\s{v}}{v} hold where 
$\Va{q} \cap Z = \Va{q} \cap W = \emptyset$. Then D-sequent \Dss{X}{F}{q}{W \cup \s{v}}{Z \cup \s{v}} holds.
\end{proposition}
\begin{mmproof}
From Lemma~\ref{lemma:holds_in_subspace} it follows that \Dss{X}{F}{q}{W}{Z} holds.
Proposition~\ref{prop:rem_bpts_increm} implies that the variables of $Z \cup \s{v}$ are redundant
in \cof{F}{q} with scope $W \cup \s{v}$. Hence  D-sequent \Dss{X}{F}{q}{W \cup \s{v}}{Z \cup \s{v}} holds.
\end{mmproof}

%
%
%
%
\section*{Proposition of Section~\ref{sec:alg_descr}: Description of \Di}
The objective of this Section is to prove the correctness of \Di~(Proposition~\ref{prop:correctness}).
To reach this objective, we need to introduce a few  new definitions and prove several lemmas.

%
%
\begin{definition}
Let \prob{X}{F} be an \ecnf formula, \pnt{q} be an assignment to \V{F} and $Z \subseteq (X \setminus \Va{q})$.
We will call D-sequent \Dss{X}{F}{q}{W}{Z} \tb{single-variable} if $|Z|$=1.
\end{definition}

%
%
\begin{definition}
D-sequents \DDs{X}{F}{q'}{W'}{v'} and  $(\prob{X}{F},\pnt{q''},W'')\!\rightarrow\!\s{v''}$
 are called \tb{compatible} if
\begin{itemize}
\item \pnt{q'} and \pnt{q''} are compatible
\item $(\Va{q'} \cup \Va{q''}) \cap (W' \cup W'' \cup \s{v'} \cup \s{v''} = \emptyset$
\end{itemize}
\end{definition}

\begin{definition}
Let \DS~ be a set of single-variable D-sequents for an \ecnf~~formula \prob{X}{F}.
We will say that \DS~is \tb{a set of compatible D-sequents} if every pair of D-sequents of \DS~is compatible.
\end{definition}
%
%
\begin{definition}
Let \DS~ be a  set of compatible D-sequents for an \ecnf~~formula \prob{X}{F}.
 Denote by  \ax~ the assignment that is the union of all  \pnt{s}  occurring in   D-sequents  $(\prob{X}{F},\pnt{s},W)$ $\rightarrow W$
of \DS. We will call \ax~the \tb{axis} of \DS. Denote by {\boldmath $W^{\DS}$} the union of the scopes $W$ of
the D-sequents of \DS.
\end{definition}

%
\begin{definition}
Let \DS~ be a set of compatible D-sequents for an \ecnf~~formula \prob{X}{F}.
Denote by {\boldmath \Xr} the set of all variables
of $X$ whose redundancy is stated by D-sequents of \DS. In the following write-up
we assume that {\boldmath $|\Xr| = |\DS|$}. That is for every variable $v$ of \Xr, set \DS~
contains exactly one D-sequent stating the redundancy of $v$. 
\end{definition}

%
%
\begin{definition}
\label{def:set_of_d_seqs}
Let \DS~ be a set of  compatible D-sequents for an \ecnf~~formula \prob{X}{F}.
 We will call D-sequent \Dss{X}{F}{\ax}{\scp}{\Xr}  \tb{the composite D-sequent} for \DS.
We will call set \DS~\tb{composable} if the composite D-sequent of \DS~holds for \prob{X}{F}.
\end{definition}%

%
%
\begin{lemma}
\label{lemma:joining_dseqs}
Let $v$ be the branching variable picked by \Di after making assignment \pnt{q}.  Assume for the sake of clarity that
$v=0$ and $v=1$ are assignments of left and right branches respectively. Denote by $\DS_0$ and $\DS_1$ the sets of D-sequents
derived in branches $v=0$ and $v=1$ respectively. Denote by \DS~the set of D-sequents produced by procedure \ti{join\_D\_seqs}
of Figure~\ref{fig:merge}. Let $\Psi$,$\Psi_0$,$\Psi_1$ be subsets of $\DS,\DS_0,\DS_1$ and 
 \Xrr{\Psi}=\Xrr{\Psi_0}=\Xrr{\Psi_1}.
Let the composite D-sequents of $\Psi_0$ and $\Psi_1$ hold. Then the composite D-sequent of $\Psi$~holds too.
\end{lemma}
\begin{mmproof}
Assume the contrary i.e.  \Dss{X}{F}{\axx{\Psi}}{\scpp{\Psi}}{\Xrr{\Psi}} does not hold.
 Then there is a $Z$-boundary point \pnt{p} of \cof{F}{\axx{\Psi}} where $Z \subseteq \Xrr{\Psi}$ that is 
\scpp{\Psi}-removable. Let $v$ be a variable of \Xrr{\Psi}. 
Denote by \pnt{q_0} and \pnt{q_1} the points $\pnt{q} \cup \s{(v=0)}$ and $\pnt{q} \cup \s{(v=1)}$
 respectively. Let \Dss{X}{F}{s_0}{W_0}{\s{v}}, \Dss{X}{F}{s_1}{W_1}{\s{v}},
 \Dss{X}{F}{s}{W}{\s{v}} be the D-sequents derived in subspaces \pnt{q_0}, \pnt{q_1} and \pnt{q} respectively. We can have
two situations here. First, all three D-sequents are equal to each other because the D-sequent of subspace \pnt{q_0} 
is symmetric in $v$. In this case, $W$=$W_0$=$W_1$. Second, the D-sequent of subspace \pnt{q} is obtained by joining
the D-sequents of subspaces \pnt{q_0} and \pnt{q_1} at variable $v$. In this case, $W = W_0 \cup W_1$.  In either
case $W_0 \subseteq W$ and $W_1 \subseteq W$ hold. Hence $\scpp{\Psi_0} \subseteq \scpp{\Psi}$ and $\scpp{\Psi_1} \subseteq \scpp{\Psi}$.

By construction, $\pnt{q_0} \subseteq \pnt{p}$ or $\pnt{q_1} \subseteq \pnt{p}$. Assume for the sake of clarity that
$\pnt{q_0} \subseteq \pnt{p}$ holds. Then point \pnt{p} is a $Z$-boundary point of \cof{F}{\axx{\Psi_0}} where $Z \subseteq \Xrr{\Psi_0}$
that is  \scpp{\Psi_0}-removable. Hence, the composite D-sequent  \Dss{X}{F}{\axx{\Psi_0}}{\scpp{\Psi_0}}{\Xrr{\Psi_0}} does not hold. 
Contradiction~$\square$
\end{mmproof}

%

%
%
\begin{lemma}
\label{lemma:subset}
Let D-sequent  \Dss{X}{F}{q}{W}{Z} hold. Let $V$ be a subset of $Z$.
Then D-sequent \Dss{X}{F}{q}{W}{V} holds too.
\end{lemma}
\begin{mmproof} 
Assume that \Dss{X}{F}{q}{W}{V} does not hold. Then there is a $V'$-boundary point \pnt{p} where $V' \subseteq V$ that is $W$-removable
in \cof{F}{q}. Since $V' \subseteq Z$ this means that $Z$ is not redundant in \prob{X}{\cof{F}{q}} with scope $W$. Contradiction.
\end{mmproof}
%
%
\begin{lemma}
\label{lemma:d_seq_blocked_var}
Let \DS~be a compatible set of  D-sequents for an \ecnf formula \prob{X}{F}.
Let \pnt{q} be an assignment to variables of \V{F} such that $\ax \subseteq \pnt{q}$ where \ax is the axis of \DS.
Let $v \in X \setminus (\Va{q}~\cup~\Xr)$ be a blocked variable of \cof{F}{q}.
Let \pnt{s} be an assignment defined as follows.
For every pair of clauses $A,B$ of $F$ that can be resolved on variable $v$,  \pnt{s} contains either
\begin{enumerate}
\item an assignment satisfying $A$ or $B$ or 
\item all the assignments of \pnt{r} such that 
  \begin{itemize}
    \item a D-sequent \DDs{X}{F}{r}{W'}{v'} is in \DS~ and 
     \item $A$ or $B$ contains variable $v'$
   \end{itemize}
\end{enumerate}
Denote by $\Psi$ the subset of \DS~comprising of all  D-sequents $(\prob{X}{F},\pnt{r})\!\!\rightarrow\!\!\s{w}$
that were used in the second condition above.
Let the composite  D-sequent \\ \Dss{X}{F}{\axx{\Psi}}{W^{\Psi}}{X^{\Psi}} hold. 
Then a D-sequent \DDs{X}{F}{\pnt{s}}{W^{\Psi} \cup \s{v}}{v} holds.
\end{lemma}
%
%
\begin{mmproof}
Notice that variable $v$ is blocked in the formula
$\cof{F}{s} \setminus (\cof{F}{s})^{X^{\Psi}}$. Then  Proposition~\ref{prop:bl_var_red} entails that $v$ is redundant
in $\cof{F}{s} \setminus (\cof{F}{s})^{X^{\Psi}}$. Since, by construction, $\axx{\Psi} \subseteq \pnt{s}$, then Lemma~\ref{lemma:holds_in_subspace}
implies that D-sequent \Dss{X}{F}{\pnt{s}}{W^{\Psi}}{X^{\Psi}} holds.
Then from Proposition~\ref{prop:rem_bpts_increm} it follows that the D-sequent \Dss{X}{F}{\pnt{s}}{\scpp{\Psi} \cup \s{v}}{X^{\Psi} \cup \s{v}} holds.
Then Lemma~\ref{lemma:subset} entails that the D-sequent \DDs{X}{F}{\pnt{s}}{\scpp{\Psi} \cup \s{v}}{v} holds $\square$
\end{mmproof}
%
%
\begin{lemma}
\label{lemma:d_seq_fls_clause}
Let \prob{X}{F} be an \ecnf. Let $C$ be a clause of $F$ falsified by an assignment \pnt{q}.
Let $v$ be a variable of $X \setminus \Va{q}$. Then D-sequent \DDs{X}{F}{\pnt{s}}{\s{v}}{v} holds
where \pnt{s} is the shortest assignment falsifying $C$.
\end{lemma}
\begin{mmproof} The proof is similar to that of Proposition~\ref{prop:unsat_clause}.
\end{mmproof}

%
%
\begin{lemma}
\label{lemma:compos_dseqs}
Any subset of active D-sequents derived by \Di is composable.
\end{lemma}
\begin{mmproof}
Let us first give an informal argument. As we mentioned in  Subsection~\ref{subsec:compos_dseqs}, 
D-sequents \DDs{X}{F}{q'}{W'}{v'} and \DDs{X}{F}{q''}{W'}{v''} may be uncomposable
if recursive reasoning is involved. That is \s{v'}-clauses are used to prove redundancy of variable $v''$ and
vice versa. \Di avoids recursive reasoning by keeping the \s{v}-clauses removed from \prob{X}{F} as long as 
a D-sequent for variable $v$ remains active.  Thus, if, for instance, \s{v'}-clauses are used to prove
redundancy of variable $v''$, the \s{v''}-clauses are removed from $F$ and cannot be used to prove
redundancy of variable $v'$.
In other words, for every path of the search tree, variables $v'$ and $v''$
are proved redundant in a particular order (but this order may be different for different paths).

Let $\Psi$ be a set of active D-sequents. To show composability of D-sequents from $\Psi$ one needs to consider the following three cases.
\begin{enumerate}
\item All D-sequents of $\Psi$ are atomic. Assume for the sake of simplicity that $\Psi = \s{S',S''}$ where 
$S'$ and $S''$ are equal to \DDs{X}{F}{q'}{W'}{v'} and \DDs{X}{F}{q''}{W'}{v''} respectively. One can have two different cases here.
\begin{itemize}
\item $S'$ and $S''$ are independent of each other. That is there is no clause $C$ of $F$ that has variables $v'$ and $v''$ and is not
 blocked at $v'$ or $v''$. In this case, one can easily show that the D-sequent \DDs{X}{F}{q' \cup q''}{W' \cup W''}{v',v''}
holds.
\item $S'$ and $S''$ are interdependent. This can happen only if $v'$ and $v''$ are blocked. Atomic D-sequents derived due to the presence
of a clause falsified by \pnt{q} (see Lemma~\ref{lemma:d_seq_fls_clause}) are independent of each other or D-sequents of blocked variables. 
Suppose the fact that  $v'$ is blocked is used to prove that  $v''$ is blocked as well.
 Then Lemma~\ref{lemma:d_seq_blocked_var} entails that $\pnt{q'} \subseteq \pnt{q''}$ and $W' \subseteq W''$ and that D-sequent
\Dss{X}{F}{q''}{W''}{Z} holds where $\s{v',v''} \subseteq Z$. Then the composability of $S'$ and $S''$ simply follows from Lemma~\ref{lemma:subset}.
\end{itemize}
\item The set $\Psi$ is obtained from set $\Psi_0$ and $\Psi_1$ when merging branches $v=0$ and $v=1$. Then Lemma~\ref{lemma:joining_dseqs}
entails that if $\Psi_0$ and $\Psi_1$ are composable, then $\Psi$ is composable as well.
\item $\Psi$ is a mix of atomic and non-atomic D-sequents. Assume for the sake of simplicity that
$\Psi = \s{S',S''}$ where $S'$ and $S''$ are equal to \DDs{X}{F}{q'}{W'}{v'} and \DDs{X}{F}{q''}{W'}{v''} respectively. Assume
that $S'$ is a result of join operations while $S''$ is atomic. Let $S'_1,\ldots,S'_k$ be the set of atomic D-sequents that are ancestors of $S'$.
Here $S'_i = \DDs{X}{F}{q'_i}{W'_i}{v'}$. Let $S''_1,\ldots,S''_k$ be the set of D-sequents obtained from $S''$ where
$S''_i = \DDs{X}{F}{q'_i \cup q''}{W''}{v''}$. Due to Lemma~\ref{lemma:holds_in_subspace}, each D-sequent $S''_i$ holds.
Since $S'_i,S''_i$ are atomic this case is covered by item 1 above and so they are composable. Then the D-sequents obtained by 
composition of $S'_i,S''_i$ can be joined producing correct D-sequents (due to correctness of operation join).  Eventually,
a correct D-sequent that is the composite of $S'$ and $S''$ will be derived $\square$
\end{enumerate}

\end{mmproof}

%
%
%
\begin{proposition}
\Di is sound and complete.
\end{proposition}
\begin{mmproof}
First, we show that \Di is \ti{complete}. \Di builds a binary search tree and
visits every node of this tree at most three times (when starting the left branch, when backtracking
to start the right branch, when backtracking from the right branch). 
So \Di is complete.

Now we prove that \Di is \ti{sound}.  
\Di terminates in two cases. First, it terminates when an empty clause is
derived, which means that $F$ is unsatisfiable. In this case, the formula $G$ returned by \Di consists
only of an empty clause.
This result is correct because this  clause is built by resolving clauses of $F$ 
and resolution is sound. Second, \Di terminates after building a sequence of D-sequents 
$(\prob{X}{F},\emptyset,X_{i_1}) \rightarrow \s{x_{i_1}}$, $\ldots$ ,$(\prob{X}{F},\emptyset,X_{i_k}) \rightarrow \s{x_{i_k}}$.
Here $x_{i_1},...,x_{i_k}$ are the variables forming $X$ and $\s{x_{i_m}} \subseteq X_{i_m} \subseteq X$, $m=1,\ldots,k$.
 We need to show that these D-sequents are correct
and composable. The latter  means that the D-sequent $(\prob{X}{F},\emptyset,X) \rightarrow X$ holds, which means that
the variables of $X$ are redundant in the current formula \prob{X}{F}.

Let us carry out the proof by induction in the number of steps of \Di\!\!.  The algorithm has two kinds of steps. A step of the first kind is
to add a new atomic D-sequent to an existing set \DS~of active D-sequents. A step of the second kind is to produce a new set of D-sequents
\DS~from the sets of D-sequents $\DS_0$ and $\DS_1$ obtained in branches $v=0$ and $v=1$.

 Let \pnt{q^k} be the assignment made by  \DS~after steps $1,\ldots,k$.
Let $\DS^k$ be the set of D-sequents maintained by \Di that are active in subspace \pnt{q^k}. (We assume here
that  every D-sequent is discarded after it takes part in a join operation. So for one redundant variable \DS~contains
only one active D-sequent.) 

The induction hypothesis is as follows. The fact that D-sequents of $\DS^k$ are individually correct
and every subset of $\DS^k$ is composable implies that the D-sequents of $\DS^{k+1}$ are correct and every subset
of $\DS^{k+1}$ is composable.

The base step, $k$=1. We need to consider the following two situations.
\begin{itemize}
\item The first atomic D-sequent $S$ is derived. In this case, its correctness follows Lemmas~\ref{lemma:d_seq_blocked_var},~\ref{lemma:d_seq_fls_clause}.
      Since $\DS^1$ consists only of one D-sequent, every subset of $\DS^1$ is obviously composable.
\item The first step consists of merging empty sets of D-sequents $\DS^1_0$ and $\DS^1_1$ derived in branches $v=0$ and $v=1$.
      In this case, \DS~is empty. So the claims that every D-sequent of \DS~is correct and all subsets are composable are vacuously true.
\end{itemize}

The induction step. We need to consider the following two situations.
\begin{itemize}
\item The set $\DS^{k+1}$ is produced by adding an atomic D-sequent $S$ to $\DS^k$. The correctness of $S$ follows
 from Lemmas~\ref{lemma:d_seq_blocked_var},~\ref{lemma:d_seq_fls_clause}. Notice that to apply Lemma~\ref{lemma:d_seq_blocked_var} we need
to use the induction hypothesis. The fact that every subset of D-sequents of $\DS^k \cup \s{S}$ is composable can be proved using the reasoning of
 Lemma~\ref{lemma:compos_dseqs}. (Notice that we cannot directly apply Lemma~\ref{lemma:compos_dseqs} because this lemma itself needs to be
proved by induction. In the sketch of a proof of Lemma~\ref{lemma:compos_dseqs}, we just gave reasoning one can use to perform such a proof.)
\item The set $\DS^{k+1}$ is produced by  merging sets of D-sequents $\DS^k_0$ and $\DS^k_1$ derived in branches $v=0$ and $v=1$.
The correctness of individual D-sequents of $\DS^{k+1}$ follows from the induction hypothesis and the 
correctness of operation join   (Proposition~\ref{prop:join_rule}).
Lemma~\ref{lemma:joining_dseqs} and the induction hypothesis entail that every subset of D-sequents of $\DS^{k+1}$ is composable.
\end{itemize}
\end{mmproof}
\vspace{5pt}

%
%
\section*{Proposition of Section~\ref{sec:compos}: Compositionality of \Di}
%
%
\begin{definition}
\label{clause_d_seq}
We will refer to D-sequents derived due to appearance of an empty clause in formula \cof{F}{q} 
(see Subsection~\ref{subsec:atomic_d_sequents}) as \tb{clause D-sequents}.
\end{definition}
%
%
\begin{proposition}[compositionality of DDS]
Let $T$ be the search tree built by \Di when solving the QE  problem
$\exists{X}[F_1 \wedge \ldots \wedge F_k]$, $\V{F_i} \cap \V{F_j}= \emptyset$, $i \neq j$. 
Let $X_i = X \cap \V{F_i}$ and $Y_i = \V{F_i} \setminus X$.
The size of  $T$ in the number of nodes
is bounded by $|\V{F}|\cdot(\eta(X_1 \cup Y_1) + \ldots +  \eta(X_k \cup Y_k))$ where 
$\eta(X_i \cup Y_i) =  2 \cdot 3^{|X_i \cup Y_i|} \cdot (|X_i| + 1), i=1,\ldots,k$
no matter how decision branching variables are chosen. 
\end{proposition}
\begin{mmproof}
Denote by $Y$ the set of variables $\V{F} \setminus X$.

We prove this proposition for a slightly modified version of \Di\!\!.  
In the version of \Di shown in Figure~\ref{fig:high_level_descr}, the D-sequents depending on the branching variable are discarded.
The modification is to keep all derived D-sequents.  This means that there 
is a set $\Pi$ where all derived D-sequents are stored. We assume that \Di does not derive the same D-sequent twice.
That is if $\Pi$ contains a D-sequent $S$ equal to \DDs{X}{F}{q}{\s{x}}{x}, then the modified \Di declares \s{x} redundant as soon as $S$ 
becomes active  instead of deriving it again.

Let $P$ be a path of $T$ and $n(v)$ be a node of $T$ that is on $P$. Here $v$ is the branching variable selected in the node $n$ by \Di\!\!.
We will call $n(v)$ a \tb{BCP node}, if  the variable $v$
was selected due to its presence in a unit clause of \cof{F}{q}. We will call $P$ an \tb{essential path}, if for every BCP node $n(v)$
lying on $P$ (if any) the latter corresponds to the \ti{right branch} of $n$. That is the variable $v$  is currently assigned
the value \ti{satisfying} the unit clause $C$ of \cof{F}{q} due to which $v$ was picked. Recall that the first value assigned to $v$ by \Di 
falsifies $C$.

Let $d$ denote the total number of nodes lying on essential paths.
Notice that the number of all nodes of $T$ is bounded by $2\cdot d$. The reason is that
a non-essential path   contains a BCP node $n(v)$ where $v$ is assigned the value falsifying the unit clause due to which $v$ was
selected.  So  the last node of this path is the left child of node $n(v)$.
 Thus the number of nodes lying only on non-essential paths is bounded by the
number of BCP nodes of $T$. Since every BCP node lies on an essential path, the total number of nodes of $T$ is bounded by $2\cdot\!d$.

Denote by \Sub{N}{ess\_paths} the total number of essential paths of $T$. Denote by \Sub{N}{res\_cl} the total number
of resolvent clauses generated by \Di\!\!. Denote  by  \Sub{N}{D\_seqs} the total number of D-sequents generated by \Di with the exception
of clause D-sequents.

 We do the rest of the proof  in two steps. First we show that 
$\Sub{N}{ess\_paths}~\leq \Sub{N}{res\_cl}~+~\Sub{N}{D\_seqs}$.
Since a path of $T$ cannot contain   more than $|X \cup Y|$ nodes, this means that the total number of nodes of $T$ is
bounded by $2\cdot |X \cup Y| \cdot (\Sub{N}{res\_cl} + \Sub{N}{D\_seqs})$. In the second step, we show that
 $2 \cdot (\Sub{N}{res\_cl} + \Sub{N}{D\_seqs}) \leq \eta(X_1 \cup Y_1) + \ldots +  \eta(X_k \cup Y_k)$ where 
$\eta(X_i \cup Y_i) =  2\cdot 3^{|X_i \cup Y_i|} \cdot (|X_i| + 1), i=1,\ldots,k$.

\vspace{3pt}
\noindent{FIRST STEP:} To prove that $\Sub{N}{ess\_paths}~\leq~\Sub{N}{res\_cl}~+~\Sub{N}{D\_seqs}$ we show that every essential path of $T$
corresponds to a new resolvent clause or a new D-sequent generated by \Di\!\! that is not a clause D-sequent.
Let $P$ be an essential path of $T$. Let $v \in X \cup Y$ be 
the first variable of $P$ picked by \Di for branching.
 The very fact that $v$ was selected 
means that some of the variables of $X$ were not proved redundant  in \prob{X}{F} yet.
Let us assume the contrary, that is \Di is able to finish $P$ without generating a new clause or a new D-sequent that is not a clause D-sequent. 
This only possible if \Di can assign all free non-redundant variables of $X$ without running into a conflict (in which case a new clause is
generated) or producing a new blocked variable  (in which case a new non-clause D-sequent  is generated). 

Let $x \in X$ be the last variable
assigned by \Di on path $P$. 
That is every other variable of $X$ is either assigned or  proved redundant before making an assignment to $x$.
Let \pnt{q} be the set of assignments on path $P$ made by \Di~before reaching  
the  node $n(x)$, and $X'$ be the set of all redundant variables of $X$ in \cof{F}{q}.
Since variables of $Y$ are assigned before  those of $X$, all non-detached  variables of $Y$ are assigned.
Then the current formula, \ie  formula $\cof{F}{q} \setminus \cf{F}{X'}{q}$ has only two kinds of clauses: 
\begin{itemize}
\item clauses depending only on detached variables of $Y$ or
\item unit clauses that depend only on variable $x$.
\end{itemize}
 The two possibilities for the unit clauses depending on $x$ are as follows.
\begin{itemize}
\item $\cof{F}{q} \setminus \cf{F}{X'}{q}$ contains both clauses $x$ and $\overline{x}$. Then, \Di 
generates a new clause. Contradiction.
\item $\cof{F}{q} \setminus \cf{F}{X'}{q}$ does not contain either $x$ or $\overline{x}$ or both. Then $x$ is blocked and
\Di generates a new non-clause D-sequent. Contradiction.
\end{itemize}

\vspace{3pt}
\noindent{SECOND STEP:} Notice that no clause produced by resolution can share variables of two different subformulas $F_i$ and $F_j$.
This means that for every clause $C$ produced by \Di\!\!, $\V{C} \subseteq (X_i \cup Y_i)$ for some $i$. The total number of clauses
depending on variables of $X_i \cup Y_i$ is $3^{|X_i \cup Y_i|}$. So  $\Sub{N}{res\_cl} \leq 3^{|X_1 \cup Y_1|} + \ldots +   3^{|X_k \cup Y_k|}$.

Now we show that $\Sub{N}{D\_seqs} \leq  |X_1|\cdot 3^{|X_1 \cup Y_1|} + \ldots +  |X_k|\cdot 3^{|X_k \cup Y_k|}$  and hence
 $2\cdot(\Sub{N}{res\_cl} + \Sub{N}{D\_seqs}) \leq \eta(X_1 \cup Y_1) + \ldots +  \eta(X_k \cup Y_k)$.
The idea is to prove that every non-clause D-sequent generated by \Di 
is \tb{limited to} \pnt{F_i}, \ie has the form  \DDs{X}{F}{s}{W}{x}
where $\Va{s} \subseteq X_i \cup Y_i$ , $W \subseteq X_i$ and $x \in  X_i$.
Recall that due to  Proposition~\ref{prop:form_replacement},  D-sequent \DDs{X}{F}{s}{W}{x} is invariant to adding  
resolvent clauses to $F$. For that reason, we will ignore the parameter \prob{X}{F} when counting
the number of D-sequents limited to $F_i$. Besides, due to Proposition~\ref{prop:increase_scope}, one can always increase
the scope of a D-sequent. For that reason, when counting D-sequents, we will also ignore the parameter $W$.
Then the total number of D-sequents limited to $F_i$ is 
equal to $|X_i| \cdot 3^{|X_i \cup Y_i|}$.  So the total number of D-sequents limited to $F_i$, $i=1,\ldots,k$ is bounded by 
$|X_1|\cdot 3^{|X_1 \cup Y_1|} + \ldots +   |X_k|\cdot 3^{|X_k \cup Y_k|}$. The factor $|X_i|$ is the number of
variables appearing on the right side of a D-sequent limited to $F_i$.  The factor $3^{|X_i \cup Y_i|}$ 
specifies the total number of all possible  assignments \pnt{s}.

Now we prove that every non-clause D-sequent derived by \Di is limited to a formula $F_i$. We carry out this proof by induction.
Our base statement is that D-sequents of an empty set are limited to $F_i$. It is vacuously true.
 Assume that the non-clause  D-sequents   generated so far are limited to $F_i$  and then show that this holds for the next
non-clause D-sequent $S$.
Let $S$ be a D-sequent \DDs{X}{F}{s}{W}{x} generated 
for a blocked variable $x \in X_i$. Such a D-sequent  is built as described in Lemma~\ref{lemma:d_seq_blocked_var}. Then \pnt{s}
consists of assignments satisfying \s{x}-clauses of $F$ or being the reason for their redundancy.  Since clauses of different subformulas
cannot be resolved with each other, every \s{x}-clause of $F$ can only have variables of $F_i$ where $x \in \V{F_i}$. 
By the induction hypothesis every non-clause D-sequent is limited to some subformula. On the other hand, \Di looks for blocked variables
when \cof{F}{q} has no empty clause. So, at the time $S$ is derived, no variable of \cof{F}{q} can be redundant due to a clause D-sequent.
This means 
that if a variable $x^*$ of an \s{x}-clause of $F$ is redundant due to D-sequent  \DDs{X}{F}{s^*}{W^*}{x^*} then $\Va{s^*} \subseteq \V{F_i}$.
So $\Va{s} \subseteq \V{F_i}$.

Now consider the case when  $S$ is obtained by joining two D-sequents $S'$, $S''$. Let us consider the following three possibilities
\begin{itemize}
\item Neither $S'$ nor $S''$ is a clause D-sequent. Then according to the induction hypothesis they should be 
limited to $F_i$. (They cannot be limited to different subformulas because then they cannot be joined due to absence of a common variable.)
Then due to  Definition~\ref{def:join_rule}, the D-sequent produced by joining $S'$ and $S''$ is also limited to $F_i$.
\item Either $S'$ or $S''$ is a clause  D-sequent. Let us assume for the sake of clarity that this is the D-sequent $S'$.
This means that $S'$ has the form (\prob{X}{F},\pnt{s},\s{x}) $\rightarrow \s{x}$
where \pnt{s} is the minimum set of assignments falsifying a clause $C$ of
$F$ and $x \in X \setminus \Va{s}$.
Since for any resolvent $C$ of $F$, $\V{C} \subseteq  \V{F_i}$, then $\Va{s} \subseteq \V{F_i}$. By the induction hypothesis,
 $S''$ is limited to $F_j$. Since $S'$ and $S''$ have at least one common variable (at which they are joined), $j$ has to be equal to $i$.
So $x \in X_i$. Then joining $S'$ with  $S''$
produces a D-sequent that is also limited to $F_i$.
\item Both $S'$ and $S''$ are clause D-sequents. We do not care about this situation because 
 by joining  $S'$ and $S''$ one obtains a clause D-sequent $\square$
\end{itemize}
\end{mmproof}



\fi

\end{document}